\documentclass[a4paper,onecolumn,11pt,accepted=2026-03-13]{quantumarticle}
\pdfoutput=1
\usepackage[utf8]{inputenc}
\usepackage[english]{babel}
\usepackage[T1]{fontenc}
\usepackage{amsmath}
\usepackage{hyperref}
\usepackage{algorithm}
\usepackage{algpseudocodex}
\usepackage{amsfonts}
\usepackage{longtable}
\usepackage{booktabs}
\usepackage[numbers]{natbib}
\usepackage{wasysym}

\usepackage{tikz}
\usepackage{lipsum}

\begin{document}

\title{Trans-dimensional Hamiltonian model selection and parameter estimation from sparse, noisy data}

\author[1,2]{Abigail N. Poteshman}
\email{poteshman@uchicago.edu}
\orcid{0000-0002-4873-4826}
\author[3]{Jiwon Yun}
\orcid{0000-0002-0959-8569}
\author[3,4]{Tim H. Taminiau}
\orcid{0000-0002-2355-727X}
\author[2,5,6]{Giulia Galli}
\email{gagalli@uchicago.edu}
\orcid{0000-0002-8001-5290}
\affil[1]{Committee on Computational and Applied Mathematics, University of Chicago, Chicago, IL 60637, USA}
\affil[2]{Materials Science Division, Argonne National Laboratory, Lemont, IL 60439, USA}
\affil[3]{QuTech, Delft University of Technology, PO Box 5046, 2600 GA Delft, The Netherlands}
\affil[4]{Kavli Institute of Nanoscience, Delft University of Technology, PO Box 5046, 2600 GA Delft, The Netherlands}
\affil[5]{Department of Chemistry, University of Chicago, Chicago, IL 60637, USA}
\affil[6]{Pritzker School of Molecular Engineering, University of Chicago, Chicago, IL 60637, USA}

\maketitle

\begin{abstract}
High-throughput characterization often requires estimating parameters and model dimension from experimental data of limited quantity and quality. Such data may result in an ill-posed inverse problem, where multiple sets of parameters and model dimensions are consistent with available data. This ill-posed regime may render traditional machine learning and deterministic methods unreliable or intractable, particularly in high-dimensional, nonlinear, and mixed continuous and discrete parameter spaces. To address these challenges, we present a Bayesian framework that hybridizes several Markov chain Monte Carlo (MCMC) sampling techniques to estimate both parameters and model dimension from sparse, noisy data. By integrating sampling for mixed continuous and discrete parameter spaces, reversible-jump MCMC to estimate model dimension, and parallel tempering to accelerate exploration of complex posteriors, our approach enables principled parameter estimation and model selection in data-limited regimes. We apply our framework to a specific ill-posed problem in quantum information science: recovering the locations and hyperfine couplings of nuclear spins surrounding a spin-defect in a semiconductor from sparse, noisy coherence data. We show that a hybridized MCMC method can recover meaningful posterior distributions over physical parameters using an order of magnitude less data than existing approaches, and we validate our results on experimental measurements. More generally, our work provides a flexible, extensible strategy for solving a broad class of ill-posed inverse problems under realistic experimental constraints.
\end{abstract}

\section{Introduction}
\label{sec:intro}

Estimation problems from sparse and noisy data are often ill-posed inverse problems, where multiple sets of parameters or models may be consistent with observed data, making unique reconstruction of ground-truth quantities nearly impossible. In such cases, the goal is not to recover a single ``true" parameter set, but to produce a posterior distribution of probable models and parameters, given a set of data.

The central motivating example for this work arises in the characterization of local spin environments in optically active spin defects in solid-state systems, such as divacancies in silicon carbide \cite{castelletto2020silicon} and nitrogen-vacancy (NV) centers in diamond \cite{rodgers2021materials}. These defects are key candidates for quantum information technologies due to their long coherence times \cite{anderson2022five, kanai2022generalized} and ability to couple coherently to nearby nuclear spins \cite{bourassa2020entanglement, bradley2022robust}. The surrounding nuclear spin bath plays a dual role: it is both the primary source of decoherence and a potential quantum resource for memory or register applications, depending on the ability to accurately resolve hyperfine couplings and spatial configurations \cite{waeber2019pulse, dong2020precise, cramer2016repeated, taminiau2012detection, taminiau2014universal}. As scalable fabrication techniques allow for the creation of billions of such spin defects \cite{marcks2024guiding, horn2024controlled}, high-throughput, data-efficient methods to characterize the nuclear spin environment of each defect are a prerequisite for the practical deployment of spin-defect-based quantum devices. Beyond characterizing the intrinsic spin environment of defects, related techniques are increasingly being applied to image external spin systems, including molecules or materials placed near a spin-defect based quantum sensor, as part of a broader effort to achieve nanoscale magnetic resonance imaging \cite{budakian2024roadmap}. 

Current approaches to estimating the parameters of the Hamiltonian describing the interaction between a central spin and neighboring nuclear spins, such as correlation spectroscopy, offer high-resolution insights but require long acquisition times and extensive manual tuning \cite{laraoui2013high, abobeih2019atomic}, making them infeasible at scale. Recently proposed machine learning methods offer some automation \cite{jung2021deep, varona2024automatic}, but rely on experimental datasets that are expensive to obtain and rely on a large amount of computing resources. Hence, a framework that can operate in the sparse, noisy data regime and still extract physically meaningful Hamiltonian parameters is desirable. Furthermore, since the number of nuclear spins interacting with the spin-defect is unknown \textit{a priori} and is itself a quantity to be estimated, the inverse problem at hand is a model dimension selection problem in addition to a Hamiltonian parameter estimation problem.

Existing approaches to similar inverse problems aim to design experimental protocols or to optimize measurement strategies to escape the ill-posed regime by collecting more informative data \cite{kura2018finite, yu2023robust, huang2023learning, anshu2021sample}. In contrast, here we assume a fixed, potentially sparse dataset, and we focus on robust inference under these constraints. Bayesian estimation techniques provide a natural language for this setting, allowing one to incorporate prior knowledge, account for measurement uncertainty, and explore the full posterior landscape of plausible models. However, performing this kind of inference efficiently—particularly in the presence of both discrete and continuous parameters, unknown model dimensionality, and high-dimensional loss landscapes riddled with local minima—remains a significant technical challenge.

To address these challenges simultaneously, we present a hybridized Markov chain Monte Carlo (MCMC) framework that combines several specialized techniques to address separately the mathematical challenges that arise. Discrete and continuous random-walk Metropolis-Hastings (RWMH) samplers are used to explore parameter spaces with both continuous and discrete parameters. Trans-dimensional model selection is handled through the reversible-jump MCMC (RJMCMC) algorithm \cite{green1995reversible}, which allows for sampling over models with differing numbers of parameters, a necessity in problems where the model dimension (e.g., the number of nuclear spins) is itself unknown. To better sample from complex and multimodal posteriors in high-dimensional spaces, we incorporate parallel tempering (PT) \cite{vousden2016dynamic}, which improves convergence and avoids trapping in local minima. Together, these methods form a generalized, flexible, and extensible framework to solve inverse problems that can operate robustly in the data-scarce regime. We demonstrate that our hybridized MCMC framework produces posterior distributions that describe simulated nuclear spin baths in the sparse, noisy data regime, and we validate our approach with experimental data to demonstrate that it achieves an accuracy comparable to much more computationally and experimentally intensive procedures, including direct machine learning approaches \cite{jung2021deep, varona2024automatic}, but using an order of magnitude fewer data points.

The remainder of this paper is organized as follows. In Sec. \ref{sec:inverse_problem}, we formalize the class of inverse problems we target and discuss the key challenges. In Sec. \ref{sec:algorithms}, we describe the MCMC algorithms and how they are hybridized to form a cohesive inference engine. In Sec. \ref{sec:hyperfine_example}, we apply the method to a specific example: recovering the structure and parameters of a spin environment from sparse coherence data. In Sec. \ref{sec:results}, we assess performance on both simulated and experimental data. Finally, in Sec. \ref{sec:discussion}, we reflect on the capabilities and limitations of this framework and suggest directions for future work.

\section{Inverse Problem}
\label{sec:inverse_problem}
\subsection{General formulation}

\begin{figure}[t]
    \centering
    \includegraphics[width=\textwidth]{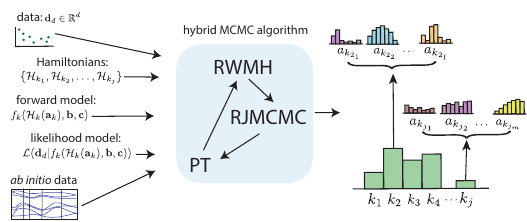} 
    \caption{\textbf{Schematic of workflow.} The input (left) to the hybrid MCMC algorithm is a fixed set of data $\mathbf{d}_d \in \mathbb{R}^d$, a family of candidate Hamiltonians $\{ \mathcal{H}_{k_1}, \mathcal{H}_{k_2}, \dots, \mathcal{H}_{k_j}\}$ which are parameterized by $\mathbf{a}_{k_i}$ where the number of parameters for different Hamiltonians $\mathcal{H}_{k_i}$ and $\mathcal{H}_{k_j}$ can be different. We also take in a forward model $f_k(\mathcal{H}_{k}(\mathbf{a}_k), \mathbf{b}, \mathbf{c})$ that generates a set of data from candidate Hamiltonian $\mathcal{H}_k$, and a likelihood model $\mathcal{L}(\mathbf{d}_d | f_k(\mathcal{H}_{k}(\mathbf{a}_k), \mathbf{b}, \mathbf{c}))$ that quantifies the probability of observing the data $\mathbf{d}_d$ given a set of model parameters, and any priors, which may come in the form of \textit{ab initio} data. The output for the hybrid MCMC algorithm is a probability distribution over the different Hamiltonian models, and for each Hamiltonian model we have a posterior distribution over the values of the parameters. }
    \label{fig:input_output_schematic}
\end{figure}

We consider some physical process that is governed by a family of Hamiltonians $\mathcal{H} \in \{\mathcal{H}_{k_1}(\mathbf{a}_{k_1}), \mathcal{H}_{k_2}(\mathbf{a}_{k_2}), \dots, \mathcal{H}_{k_j}(\mathbf{a}_{k_j}) \}$, where each Hamiltonian $\mathcal{H}_{k_i}$ is parametrized by $k_i$ variables given by $\mathbf{a}_{k_i} = (a_{k_{1}}, a_{k_{2}}, \dots, a_{k_i})$. We consider a sparse set of data $\mathbf{d}_d \in \mathbb{R}^d$ that is generated by a forward model $\mathbf{d}_d = f(\mathcal{H}_k(\mathbf{a}_k), \mathbf{b}, \mathbf{c}) + \mathbf{\epsilon}_d$ where the forward model $f(\cdot)$ depends on the parameterized Hamiltonian $\mathcal{H}_k(\mathbf{a}_k)$, unknown parameters $\mathbf{b}$, known parameters $\mathbf{c}$, and $\epsilon_d$ is a noise model. Examples of unknown parameters represented by $\mathbf{b}$ include parameters that are not directly captured by Hamiltonian terms but are nevertheless crucial to reproduced experimental data, such as experimental setup-dependent calibration and sensitivity effects or emergent effects that may not be directly accounted for in the Hamiltonian. Examples of known parameters modeled by $\textbf{c}$ may include known experimental settings, such as magnetic field or temperature. Formally, the inverse problem we seek to resolve is: 
\begin{align}
    \text{arg}\min_{k, \mathbf{a}_k, \mathbf{b}} || f(\mathcal{H}_k(\mathbf{a}_k)), \mathbf{b}, \mathbf{c}) - \mathbf{d}_d||_{N(\mathbf{\epsilon}_d)}.
\end{align} That is, we seek to find the candidate model $\mathcal{H}_k$, indexed by $k$, and the parameters $\mathbf{a}_k$ and $\mathbf{b}$ that minimize the norm of the residual between some forward model $f$ applied to the Hamiltonian of dimension $k$ with parameters $\mathbf{a}_k$ and $\mathbf{b}$ and some $d$-dimensional data, where the norm $N(\epsilon_d)$ depends on underlying assumptions on the noise model $\epsilon_d$. The assumptions on the noise model impact the appropriate likelihood function $\mathcal{L} = \mathbb{P}(\mathbf{d}|\mathcal{H}_k(\mathbf{a}_k), \mathbf{b}, \mathbf{c})$, which is the probability of observing data $d$ given $\mathcal{H}_k(\mathbf{a}_k), \mathbf{b},$ and $\mathbf{c}$. Without loss of generality, we assume that the dimension of $\mathbf{a}_k$ is variable and the dimension of $\mathbf{b}$ is fixed, but we note that the same strategies that are applied to account for the variable number of dimensions of $\mathbf{a}_k$ may be applied to $\mathbf{b}$.

Furthermore, we acknowledge that ``sparse" and ``noisy" are quantities that are context-dependent, and we generally consider data $\mathbf{d}_d$ to be ``sparse" and ``noisy" if there are multiple parameterizations $k_i, k_j$ and $(\mathbf{a}_{k_i}, \mathbf{b}_i), (\mathbf{a}_{k_j}, \mathbf{b}_j)$ such that 
\begin{align}
    \min_{k_i, \mathbf{a}_{k_i}} || f(\mathcal{H}_{k_i}(\mathbf{a}_{k_i}), \mathbf{b}_i, \mathbf{c}) - \mathbf{d}_d||_{N(\mathbf{\epsilon}_d)} &\approx \min_{k_j, \mathbf{a}_{k_j}} || f(\mathcal{H}_{k_j}(\mathbf{a}_{k_j}), \mathbf{b}_j, \mathbf{c}) - \mathbf{d}_d||_{N(\mathbf{\epsilon}_d)}
\end{align} for a fixed set of data $\mathbf{d}_d$. That is, the sparsity and noise regimes we consider here are with respect to the ill-posedness of a specific inverse problem.

Since we will focus on the regime where the inverse problem is ill-posed, we take a Bayesian approach, as we know that we will not be able to exactly recover $k$, $\mathbf{a}_k$, and $\mathbf{b}$, but we will instead recover posterior distributions over $k$, $\mathbf{a}_k$, and $\mathbf{b}$. That is, instead of a single $k$ and $\mathbf{a}_k$, we will recover a set of $\{k_l, k_m, \dots, k_n \}$ with corresponding probabilities $\{p_{k_l}, p_{k_m}, \dots, p_{k_n} \}$ (where $\sum_j p_{k_j} = 1$), with the interpretation the data $\mathbf{d}_d$ is generated by $\mathcal{H}_{k_i}$ with probability $p_i$. For each model class $k_i$, we will also obtain posterior distributions over the parameters $\{\mathbf{a}_{k_{i_1}}, \mathbf{a}_{k_{i_2}}, \dots \mathbf{a}_{k_{i_p}}\}$ with probabilities $\{p_{k_{i_1}}, p_{k_{i_2}}, \dots, p_{k_{i_p}}\}$ and $\sum_j p_{k_{i_j}} = 1$, and likewise for $\mathbf{b}$ (see Fig. \ref{fig:input_output_schematic}).

\subsection{Hybridized MCMC-based Hamiltonian model selection and parameter estimation}

Markov chain Monte Carlo (MCMC) algorithms provide a robust framework for exploring complex, high-dimensional parameter spaces by generating a sequence of samples that approximate a target probability distribution. In general, these algorithms initialize a set of walkers across the domain of a parameter space and propose updates to the parameters based on a specified transition rule. The acceptance or rejection of a proposed move is governed by whether the likelihood (i.e., the probability of observing some data given a set of parameters) increases for an updated set of parameters. After an initial burn-in period, the sequence of walker positions represents samples from the posterior distribution, which provides the most probable parameter values given the observed data.

As shown in Fig. \ref{fig:input_output_schematic}, the hybridized MCMC algorithm takes as input a fixed dataset $\mathbf{d}_d \in \mathbb{R}^d$, a set of candidate Hamiltonian models $\{ \mathcal{H}_{k_1}, \mathcal{H}_{k_2}, \dots, \mathcal{H}_{k_j} \}$, each parameterized by a potentially different number of parameters $\mathbf{a}_{k_i}$, and a forward model $f_k(\mathcal{H}_{k}(\mathbf{a}_k), \mathbf{b}, \mathbf{c})$ that simulates data for a given Hamiltonian. The algorithm also requires a likelihood function $\mathcal{L}(\mathbf{d}_d \mid f_k(\mathcal{H}_{k}(\mathbf{a}_k), \mathbf{b}, \mathbf{c}))$ to quantify how well the simulated data from a candidate model match the measured data, as well as any available prior information, such as constraints from \textit{ab initio} calculations. The algorithm produces as outputs posterior distributions over the candidate Hamiltonian models and, for each model, a posterior distribution over its associated parameter values.

We address the challenge of recovering parameters that may be drawn from different domains (i.e., the set or distribution from which parameters are sampled) by partitioning the parameter space into subsets, $(\mathbf{a}_k, \mathbf{b}) = (a_1, \dots, a_{i} | a_{i+1}, \dots, a_{k} | b_1, \dots, b_j)$, where different partitions can follow either discrete distributions based on precomputed or experimental datasets or continuous distributions suited for a broader search, and subsets of variables can also have constant or varying dimensions. This partitioning enables efficient sampling by tailoring MCMC strategies to different types of parameters. The hybrid algorithm includes different MCMC strategies appropriate for each subset of parameters. Since the approach accommodates arbitrary likelihood functions and arbitrary forward models (see Fig. \ref{fig:input_output_schematic}), our framework is designed for extensibility across a variety of physical systems and experimental setups; although a few key MCMC algorithms for dealing with varying dimensionality and efficient exploration of non-linear, non-convex landscapes have already been implemented, additional MCMC-based algorithms to evolve subsets of given parameters may be combined and implemented. 

\section{Algorithms}
\label{sec:algorithms}

\begin{figure}[t]
    \centering
    \includegraphics[width=\textwidth]{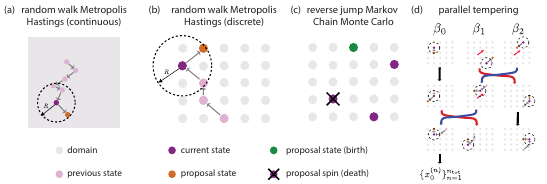} 
    \caption{\textbf{Schematic of algorithms} (a) random walk Metropolis Hastings in over a continuous domain (Alg. \ref{alg:rwmh}), (b) random walk Metropolis Hastings over a discrete domain (Alg. \ref{alg:rwmh}), (c) reverse jump Markov chain Monte Carlo (Alg. \ref{alg:rjmcmc}), and (d) parallel tempering (Alg. \ref{alg:parallel_temp}).}
    \label{fig:algorithm_schematic}
\end{figure}

In this section, we elaborate on details of the implementation of the hybrid algorithm and provide pseudocode for the sub-algorithms of the hybridized MCMC procedure (Fig. \ref{fig:algorithm_schematic}). For details on these Monte-Carlo based methods, we refer the reader to \cite{denison2002bayesian, sanz2024first}. While many of these algorithms are standard within the MCMC literature, we present them here with consistent notation to demonstrate how they can be hybridized with each other for different types of continuous and discrete parameters and how they can be used for specific sampling tasks.

\subsection{Random walk Metropolis-Hastings (RWMH) for continuous and discrete domains}

\begin{algorithm}[t]
    \caption{Random walk Metropolis Hastings (RWMH)}
    \label{alg:rwmh}
    \begin{algorithmic}[1]
        \Statex \textbf{Input}: Data $\mathbf{d}_d$, partition of a fixed-length set of parameters to update $\mathbf{p} = \{p_1, \dots, p_m \}$ and fixed parameters $\mathbf{q} = \{q_1, \dots, q_s\}$, with $\mathbf{p} \cup \mathbf{q} = \{a_1, \dots, a_k, b_1, \dots, b_j \}$ and $\mathbf{p} \cap\mathbf{q} = \emptyset$ and their initial values $\mathbf{p}^{(0)}= (p_1^{(0)}, p_2^{(0)}, \dots, p_m^{(0)})$ and $\mathbf{q}^{(0)}= (q_1^{(0)}, q_2^{(0)}, \dots, q_s^{(0)})$, known parameters $\mathbf{c}$, fixed Hamiltonian model $\mathcal{H}_k(\mathbf{a}_k)$ indexed by $k$, forward model $f_{k}(\mathcal{H}_{k}(\mathbf{a}_{k}), \mathbf{b}, \mathbf{c}) = f_k(\mathbf{p}, \mathbf{q}, \mathbf{c})$, likelihood model $\mathcal{L}(\mathbf{d}_d|f_k(\mathbf{p},\mathbf{q},\mathbf{c}))$, domain $\mathcal{D}$, proposal kernel $r_{R, \mathcal{D}}(x, z)$, radius $R$, number of steps $N_{\text{step}}$  
        
        \For{$n = 0, \dots N_{\text{step}}-1$}
            \For {$l = 1, \dots, m$}
            \State \textbf{Proposal step}: Sample $p^*_{l} \sim r_{R, \mathcal{D}}(p_{l}^{(n)}, \cdot)$; \\ $\mathbf{p}^* = \{p^{(n)}_1, \dots, p^{(n)}_{l-1}, p^*_l, p^{(n)}_{l+1}, \dots, p^{(n)}_m\}$
            \State \textbf{Accept/reject step}: Update 
            \begin{align*}
                \mathbf{p}^{(n+1)} &= 
                \begin{cases}
                \mathbf{p}^* & \text{with probability } \alpha(\mathbf{p}^{(n)}, \mathbf{p}^*) = \min\{1, \frac{\mathcal{L}(\mathbf{d}_d|f_k(\mathbf{p}^*, \mathbf{q}, \mathbf{c}))}{\mathcal{L}(\mathbf{d}_d|f_k(\mathbf{p}^{(n)}, \mathbf{q}, \mathbf{c}))} \frac{r_{R, \mathcal{D}}(p^*_l, p^{(n)}_l)}{r_{R, \mathcal{D}}(p^{(n)}_l, p^*_l)} \}\\
                \mathbf{p}^{(n)} & \text{with probability } 1 - \alpha(\mathbf{p}^{(n)}, \mathbf{p}^*)
                \end{cases} \\
                \mathbf{q}^{(n+1)} &= \mathbf{q}^{(n)}
            \end{align*}
            \EndFor
        \EndFor
        \Statex \textbf{Output}: Sequence of updated parameters: $ \{\mathbf{p}^{(n)} \}_{n=1}^{N_{\text{step}}} = (\mathbf{p}^{(1)}, \mathbf{p}^{(2)}, \dots \mathbf{p}^{(N_{\text{step}})})$ and sequence of fixed parameters: $\{\mathbf{q}^{(n)} \}_{n=1}^{N_{\text{step}}} = (\mathbf{q}^{(1)}, \mathbf{q}^{(2)}, \dots \mathbf{q}^{(N_{\text{step}})}) = (\mathbf{q}^{(0)}, \mathbf{q}^{(0)}, \dots, \mathbf{q}^{(0)})$    \end{algorithmic}
\end{algorithm}

We employ a random walk Metropolis-Hastings (RWMH) algorithm to sample a subset of unknown parameters with fixed dimension, allowing for both continuous and discrete domains (see Alg. \ref{alg:rwmh}). Specifically, we partition the set of unknown parameters into $\mathbf{p} = \{p_1, \dots, p_m\}$, which are updated during the sampling, and $\mathbf{q} = \{q_1, \dots, q_s\}$, which remain fixed. At each iteration, a single parameter $p_l^*$ is proposed from a domain-aware proposal distribution $r_{R, \mathcal{D}}(p_l^{(n)}, \cdot)$, where the domain $\mathcal{D}$ and the radius $R$ define the step size and allowable transitions. The proposal mechanism is flexible, permitting updates in either a discrete or continuous setting, depending on the nature of the domain $\mathcal{D}$ from which the parameters are sampled (see Fig. \ref{fig:algorithm_schematic}a \& b). For continuous variables, this can be a closed set over a fixed interval, in which case there is a standard interpretation of a random walk. For a discrete domain $\mathcal{D}$, there may be cases, such as selecting different configurations of spins over lattice sites that have real-space coordinates, where there is still a standard interpretation of a random walk in real space. For discrete sets without parameters that have a natural mapping to real space, such as possible energy transitions or interaction strengths, one may need a problem-specific interpretation of a random walk to use RWMH. The proposed state $\mathbf{p}^*$ (containing $p_l^*$) is then accepted or rejected based on the Metropolis criterion, ensuring detailed balance. The likelihood function $\mathcal{L}$ governs the acceptance probability, leveraging a forward model $f(\mathbf{p}, \mathbf{q}, \mathbf{c})$ to compare simulated and observed data. Although this is a standard MCMC algorithm, we emphasize the practical importance of the built-in partitioning of parameters in the implementation described in Alg. \ref{alg:rwmh}; such partitioning is particularly relevant in a Bayesian approach to Hamiltonian parameter estimation problems for which prior information on certain parameters may be available from \textit{ab initio} calculations or experimental data.

\subsection{Reverse Jump Markov Chain Monte Carlo (RJMCMC)}
\label{subsec:rjmcmc}

In the previous algorithm (RWMH), the parameters are partitioned into a set to be updated and a set to remain fixed, but the dimensions of the updated set are fixed. To account for scenarios in which the number of parameters must be estimated in addition to the parameters themselves, we use a MCMC-based technique known as ``reverse jump" Markov chain Monte Carlo, which is a method for trans-dimensional Bayesian model selection (see Alg. \ref{alg:rjmcmc}) \cite{green2009reversible}. The RJMCMC method for trans-dimensional model selection has been applied to inverse problems across the physical and biological sciences, including estimating and studying the number of black holes from gravitational wave data \cite{toubiana2023there, zevin2017constraining}, seismic inversion problems that arise in geophysics \cite{zhu2018seismic, cho2018quasi}, and inferring phylogenetic trees from genomic data \cite{oaks2022generalizing, pagel2008modelling}. This approach enables simultaneous trans-dimensional model selection and parameter estimation, allowing for a systematic exploration of the parameter space in cases where standard regularization techniques, such as sparsity-inducing regularization, may not be physically meaningful.

When using RJMCMC, we partition the set of parameters in $\mathbf{p}$ and $\mathbf{q}$, where $\mathbf{p}$ is a set of parameters to be estimated using RJMCMC where the number of parameters $\mathbf{p}$ is unknown, and $\mathbf{q}$ is a set of unknown parameters that remains fixed. RJMCMC extends traditional MCMC by allowing proposals between models of different dimensionality, enabling the simultaneous exploration of both parameter values and model structures through dimension-changing proposal moves that ensure detailed balance and reversibility. In the discrete case, RJMCMC operates as a birth-death process where new parameters (e.g., additional particles or interactions) are ``born" into the model or ``die" based on probabilistic moves that maintain detailed balance (see Fig. \ref{fig:algorithm_schematic}c) \cite{denison2002bayesian}. In the continuous case, dimension matching is ensured by introducing auxiliary variables that transform proposals between parameter spaces of different dimensions, preserving reversibility by maintaining a one-to-one mapping with a Jacobian determinant correction \cite{green1995reversible}.

\begin{algorithm}[t!]
    \caption{Reverse Jump Markov Chain Monte Carlo (RJMCMC) for discrete parameters}
    \label{alg:rjmcmc}
    \begin{algorithmic}[1]

        \Statex \textbf{Input}: Data $\mathbf{d}_d$, partition of variable parameters with variable length $\mathbf{p} = \{p_1, \dots, p_m \}$ and fixed parameters with fixed length $\mathbf{q} = \{q_1, \dots, q_s\}$ and $\mathbf{p} \cup \mathbf{q} = \{a_1, \dots, a_k, b_1, \dots, b_j \}$ and $\mathbf{p} \cap\mathbf{q} = \emptyset$, known parameters $\mathbf{c}$, family of Hamiltonians $\{\mathcal{H}_k \}_{k=1}^{k_{\text{max}}}$, family of forward models $\{f_k(\mathcal{H}_k(\mathbf{a}_k), \mathbf{b}, \mathbf{c}) \}_{k=1}^{k_{\text{max}}}$, with $f_{k}(\mathcal{H}_{k}(\mathbf{a}_{k}), \mathbf{b}, \mathbf{c}) = f_k(\mathbf{p}, \mathbf{q}, \mathbf{c})$, initial values $\mathbf{p}^{(0)}= (p_1^{(0)}, p_2^{(0)}, \dots, p_m^{(0)})$, $\mathbf{q}^{(0)}= (q_1^{(0)}, q_2^{(0)}, \dots, q_s^{(0)})$, $k^{(0)}$, likelihood model $\mathcal{L}
            (\mathbf{d}_d|f_k(\mathbf{p},\mathbf{q},\mathbf{c}))$, domain $\mathcal{D}$, number of steps $N_{\text{step}}$, dimension-changing kernel $\gamma(k, \cdot)$
       
        \For{$n = 0, \dots N_{\text{step}}-1$}
        \State \textbf{Proposal step}: sample $k^* \sim \gamma(k^{(n)}, \cdot)$
        \If{$k^* > k^{(n)}$}
        \begin{align*}
            k^* &= k^{(n)}+1\\
            p^*_{k+1} &\sim \mathcal{D}\\
            \mathbf{p}^* &= (p_1^{(n)}, p_2^{(n)}, \dots, p_k^{(n)}, p_{k+1}^*)
        \end{align*}
        \Else
        \begin{align*}
            k^* &= k^{(n)}-1\\
            l &\sim \{1, 2, \dots,k^{(n)} \} \\
            \mathbf{p}^* &= (p_1^{(n)}, \dots, p_{l-1}^{(n)}, p_{l+1}^{(n)}, \dots, p_k^{(n)})
        \end{align*}
        \EndIf

        \State \textbf{Accept/reject step}: Update 
            \begin{align*}
                k^{n+1}, \mathbf{p}^{(n+1)} &= 
                \begin{cases}
                k^*, \mathbf{p}^* & \text{with probability } \alpha(k^{(n)},\mathbf{p}^{(n)}; k^*, \mathbf{p}^*) \\
                k^{(n)}, \mathbf{p}^{(n)} & \text{with probability } 1 - \alpha(k^{(n)},\mathbf{p}^{(n)}; k^*, \mathbf{p}^*)
                \end{cases} \\
                \mathbf{q}^{(n+1)} &= \mathbf{q}^{(n)} \\
                \alpha(k^{(n)},\mathbf{p}^{(n)}; k^*, \mathbf{p}^*) &= \min\{1, \frac{\mathcal{L}(\mathbf{d}_d|f_{k^*}(\mathbf{p}^*, \mathbf{q}, \mathbf{c})}{\mathcal{L}(\mathbf{d}_d|f_{k^{(n)}}(\mathbf{p}^{(n)}, \mathbf{q}, \mathbf{c})} \frac{\gamma(k^*, k^{(n)})}{\gamma(k^{(n)}, k^*)} \}
            \end{align*}
    
        \EndFor
        \Statex \textbf{Output}: $\{ k^{(n)}\}_{n=1}^{N_{\text{step}}}$, $\{ \mathbf{p}^{(n)}\}_{n=1}^{N_{\text{step}}}$, $\{ \mathbf{q}^{(n)}\}_{n=1}^{N_{\text{step}}}$
    \end{algorithmic}
\end{algorithm}

\subsection{Parallel Tempering (PT)}
\label{subsec:pt}
Parallel tempering, also known as replica exchange Monte Carlo, is a computational technique used to efficiently sample complex energy landscapes, particularly in systems with many local minima. The method involves running multiple copies (replicas) of the system in parallel, each at a different temperature. Higher temperature replicas explore the energy landscape more freely, overcoming barriers between local minima, while lower temperature replicas focus on finding the global minimum. At regular intervals, configurations between replicas are exchanged with a probability determined by the Metropolis criterion (see Fig. \ref{fig:algorithm_schematic}d). This exchange allows for low-temperature replicas to escape local minima by occasionally adopting the high-temperature configuration. The exchanges are designed to maintain detailed balance, ensuring that the sampling process correctly represents the equilibrium distribution at each temperature. By allowing configurations to move between different temperature levels, parallel tempering enhances the ability to explore the energy landscape, reducing the risk of getting trapped in local minima. This leads to more efficient and accurate sampling, particularly for systems with rugged energy landscapes. PT is an acceleration technique for a fixed number of parameters in highly non-linear non-convex energy landscapes.

\begin{algorithm}[t]
    \caption{Parallel tempering (PT)}
    \label{alg:parallel_temp}
    \begin{algorithmic}[1]
        \Statex \textbf{Input}: Number of strands $J$, inverse temperatures $\beta_1 = 1 > \dots > \beta_J$, data $\mathbf{d}_d$, partition of a fixed-length set of parameters to update $\mathbf{p} = \{p_1, \dots, p_m \}$ and fixed parameters $\mathbf{q} = \{q_1, \dots, q_s\}$, with $\mathbf{p} \cup \mathbf{q} = \{a_1, \dots, a_k, b_1, \dots, b_j \}$ and $\mathbf{p} \cap\mathbf{q} = \emptyset$ and their initial values $\mathbf{p}^{(0)}= (p_1^{(0)}, p_2^{(0)}, \dots, p_m^{(0)})$ and $\mathbf{q}^{(0)}= (q_1^{(0)}, q_2^{(0)}, \dots, q_s^{(0)})$, known parameters $\mathbf{c}$, fixed Hamiltonian model $\mathcal{H}_k(\mathbf{a}_k)$ indexed by $k$, likelihood model $\mathcal{L}(\mathbf{d}_d|f_k(\mathbf{a_k}, \mathbf{b}, \mathbf{c}))$, domain $\mathcal{D}$, number of steps $N_{\text{step}}$

        \For{$n = 0, \dots N_{\text{step}}-1$}
        \State \textbf{Within-strand step}: Generate $\Tilde{\mathbf{p}}_j^{(n+1)}$ from $\mathbf{p}_j^{(n+1)}$ by one step of (Alg. \ref{alg:rwmh}) for each strand $j \in [1, \dots, J ]$ using the corresponding likelihood model 
        \begin{align*}
            \mathcal{L}_j(\mathbf{d}_d|f_k(\mathbf{p}, \mathbf{q},\mathbf{c})) = (\mathcal{L}(\mathbf{d}_d|f_k(\mathbf{p}, \mathbf{q},\mathbf{c})))^{\beta_j}
        \end{align*}
        \State \textbf{Draw strands to attempt swap}: Sample $a, b \sim \{1, \dots J\}$ with $a \neq b$ uniformly at random
        \State \textbf{Evolve non-swapped chains forward}: For $j \notin \{a, b\}$, set 
        \begin{align*}
            \mathbf{p}_j^{(n+1)}, \mathbf{q}_j^{(n+1)} = \Tilde{\mathbf{p}}_j^{(n+1)}, \mathbf{q}_j^{(n)}
        \end{align*}
        \State \textbf{Swap proposal step}: Attempt a swap of states between strands $a$ and $b$:
        \begin{align*}
            (\mathbf{p}_a^{(n+1)}, \mathbf{p}_b^{(n+1)}) &= \begin{cases}
                (\tilde{\mathbf{p}}_b^{(n+1)}, \tilde{\mathbf{p}}_a^{(n+1)}) &\text{with probability } \alpha_{\text{PT}}(a, b)\\
                 (\tilde{\mathbf{p}}_a^{(n+1)}, \tilde{\mathbf{p}}_b^{(n+1)}) &\text{with probability } 1 - \alpha_{\text{PT}}(a, b)
            \end{cases} \\
            \mathbf{q}_a^{(n+1)}, \mathbf{q}_b^{(n+1)} &= \mathbf{q}_a^{(n)}, \mathbf{q}_b^{(n)}\\
           \alpha_{\text{PT}} &= \min \{1, \frac{\mathcal{L}_a(\mathbf{d}_d|\tilde{\mathbf{p}}_b^{(n+1}, \mathbf{q}, \mathbf{c})}{\mathcal{L}_a(\mathbf{d}_d|\tilde{\mathbf{p}}_a^{(n+1}, \mathbf{q}, \mathbf{c})} \frac{\mathcal{L}_b(\mathbf{d}_d|\tilde{\mathbf{p}}_a^{(n+1)}, \mathbf{q}, \mathbf{c})}{\mathcal{L}_b(\mathbf{d}_d|\tilde{\mathbf{p}}_b^{(n+1)}, \mathbf{q}, \mathbf{c})} \}
        \end{align*}
        \EndFor
        \Statex \textbf{Output}: Sequences corresponding to the $j=1, \beta_j=1$ chain: $\{ \mathbf{p}_{j=1}^{(n)}\}_{n=1}^{N_{\text{step}}}$, $\{ \mathbf{q}_{j=1}^{(n)}\}_{n=1}^{N_{\text{step}}}$
    \end{algorithmic}
\end{algorithm}

\subsection{Hybrid algorithm}
\label{subsec:hybrid_alg}

Hybridizing Markov chain Monte Carlo (MCMC) algorithms can enhance sampling efficiency and model selection for complex inverse problems. A robust approach combines the random walk Metropolis-Hastings (RWMH) algorithm for exploring continuous parameter spaces with reversible jump MCMC (RJMCMC) for trans-dimensional moves that enable discrete model selection. To improve within-model fits, particularly in highly non-linear and non-convex likelihood landscapes, parallel tempering (PT) based on RWMH can be incorporated for a fixed number of discrete parameters, allowing for enhanced exploration of local minima and better posterior sampling. This multi-tiered strategy balances global model selection with efficient local exploration, mitigating issues like poor mixing and mode trapping, which are common in high-dimensional and multi-modal inverse problems.

\section{Example application: Recovering hyperfine coupling parameters of a nuclear spin bath of defects in semiconductors from coherence data}
\label{sec:hyperfine_example}

While the trans-dimensional Hamiltonian model selection and parameter estimation problem in the sparse, noisy regime is general, we validate our implementation and illustrate the utility of a hybrid MCMC-based approach on a specific characterization problem of interest to the quantum information community: the recovery of hyperfine couplings of a nuclear spin bath of a single spin defect in a semiconductor from sparse, noisy coherence data. This example problem is a trans-dimensional Hamiltonian selection problem, since the number of nuclear spins in the bath is not known \textit{a priori}; in addition, it is a problem for which there is interest in accelerating recovery of spin environments for purposes of device scale-up when using spin defects in seminconductors as units of quantum information processing, with nuclear spins as auxiliary qubits for quantum algorithms and memory. This problem further illustrates the ability of a hybrid MCMC approach to deal with parameters for which strong priors via \textit{ab initio} calculations can be incorporated, as well as parameters that are drawn from a continuous distribution. 

\subsection{Hamiltonian and forward models}
We consider a  nuclear spin bath consisting of $^{13}$C surrounding a nitrogen vacancy (NV) center in diamond. We seek to characterize the number of spins in the $^{13}$C nuclear spin bath and the strength of each spin's hyperfine coupling $(A_{\perp}, A_{\parallel})$. We consider the Hamiltonians indexed by the variable of the number of nuclear spins $k$ with the Hamiltonian given by:
\begin{align}
    \mathcal{H}_k = D\mathcal{S}_z^2 + \gamma_e B_z \mathcal{S}_z + \sum_{i=1}^k \gamma_n B_z \mathcal{I}_{z,i} + \sum_{i=1}^k \mathcal{S}_z(A_{\parallel,i}\mathcal{I}_{z,i} + A_{\perp,i}\mathcal{I}_{x,i})
\end{align} where the known quantities are $D = 2\pi \times2.87$ GHz, $B_z$ is the strength of an externally applied magnetic field along the axis of the NV center, $\gamma_{e}$ is the gyromagnetic ratio of an electron, $\gamma_n$ is the nuclear gyromagnetic ratio of the nuclei, and $\mathbf{\mathcal{S}} = (\mathcal{S}_x, \mathcal{S}_y, \mathcal{S}_z)$ and $\mathbf{\mathcal{I}} = (\mathcal{I}_x, \mathcal{I}_y, \mathcal{I}_z)$ are the spin operators for the electronic and nuclear spins, respectively. We neglect nuclear–nuclear interactions in the Hamiltonian, as we expect these interactions to be negligible for the relatively short $\tau$ (< 10 $\mu$s) and relatively low pulse number ($\leq$ 32 CP pulses) in the experiments we consider for high-throughput characterization. We seek to recover the number of spins $k$, which determines the dimension of the Hamiltonian, and for each spin, we aim to recover the parallel and perpendicular components ($A_{\parallel}, A_{\perp}$) of the hyperfine coupling tensor, which characterizes the coupling strength between the nuclei and the NV center.

The forward model $f_k(\mathcal{H}_k(\mathbf{a}_k), \mathbf{b}, \mathbf{c})$ for a set of $k$ nuclear spins with hyperfine components $\mathbf{A_{\parallel}}, \mathbf{A_{\perp}}$ under dynamical decoupling experiment with $N$ CP-pulses measured at inter-pulse spacing $\tau_j$ in external applied magnetic field $B_z$ is:
\begin{align}
f_k(\mathcal{H}_k(\mathbf{A_{\parallel}}, \mathbf{A_{\perp}}, B_z), \lambda, N, \tau_j) &= (\frac{1}{2}(1 + \prod_{i=1}^{k} M_i(A_{\parallel, i}, A_{\perp, i}, N, B_z, \tau_j)))^{-\frac{\tau_j}{\lambda}} \\
M_i(A_{\parallel, i}, A_{\perp, i}, N, B_z, \tau) &= (1 - m_{i,x}^2 \frac{(1- \cos{\alpha_i})(1-\cos{\beta})}{1 + \cos{\alpha_i}\cos{\beta} - m_{i,z}\sin{\alpha_i}\sin{\beta}} \sin{\frac{N\phi_i}{2}}^2) \\
    \cos{\phi_i} &= \cos{\alpha_i}\cos{\beta} - m_{i,z}\sin{\alpha_i}\sin{\beta} \\
    m_{i,z} &= \frac{A_{\parallel, i} + \omega_L}{\tilde{\omega}_i} \\
    m_{i,x} &= \frac{A_{\perp, i}}{\tilde{\omega}_i} \\
    \tilde{\omega}_i &= \sqrt{(A_{\parallel, i} + \omega_L)^2 + A_{\perp, i}^2} \\
    \alpha_i &= \tilde{\omega}_i \tau_j \\
    \beta &= \omega_L \tau_j \\
    \omega_L &= - \gamma_n B_z,
\end{align} where we can consider our partition of parameters to be $\mathbf{a}_k = \{\mathbf{A}_{\parallel}, \mathbf{A}_{\perp} \}$, $\mathbf{b} = \{ \lambda\}$, and $\mathbf{c} = \{N, B_z, \tau_j\}$ \cite{taminiau2012detection}.

\subsection{Likelihood model and priors}
Given that coherence signals are measured with photon counts on the order of thousands to tens of thousands, we approximate the noise as Gaussian rather than Poisson-distributed. Accordingly, we define the likelihood function as:

\begin{align}
\label{eq:l2_likelihood}
\mathcal{L}(\mathbf{d}_d|f_k(\mathcal{H}_k(\mathbf{a_k}), \mathbf{b}, \mathbf{c})) &= \exp\left(-\frac{1}{2 \sigma^2} \sum_{i=1}^d(\mathbf{d}_i - f_k(\mathcal{H}_k(\mathbf{A_{\parallel}}, \mathbf{A_{\perp}}, B_z), \lambda, N, \tau_i))^2\right),
\end{align}
where $\sigma^2$ is a hyperparameter that relates to assumptions on the variance of the noise. The partition of parameters is $\mathbf{a}_k = (A_{\parallel, 1}, A_{\perp, 1}, A_{\parallel, 2}, A_{\perp, 2}, \dots, A_{\parallel, k}, A_{\perp, k})$, $\mathbf{b} = (\lambda)$, and $\mathbf{c} = (N, \tau_j, B_z)$. The set $\mathbf{a}_k$ corresponds to the hyperfine parameters of the nuclear spins in the bath; since we do not know \textit{a priori} how many nuclear spins are present, we do not know \textit{a priori} how many hyperfine parameters $(A_{\parallel}, A_{\perp})$ we need to recover. The continuous parameter $\mathbf{b} = (\lambda)$ that we restrict to the interval $[0, 1 ]$ accounts for relative decoherence effects. This parameter is crucial for experimental consistency and likely reflects interactions between the NV center and unaccounted lattice impurities, such as nitrogen substitutional defects, and the broader, distant spin bath, which causes and overall decoherence decay. The known parameters $\mathbf{c} = (N, B_z, \tau_j)$ are the externally applied magnetic field $B_z$, the number of laser pulses $N$ required in the experimental protocol to obtain a coherence signal, and the amount of time $\tau_j$ in between sequential laser pulses. 

To enhance the efficiency of parameter estimation and model selection, we integrate data from \textit{ab initio} calculations as physical constraints. Specifically, we restrict possible nuclear positions to specific lattice sites of diamond, leveraging computational predictions of hyperfine couplings at these sites based on the Fermi contact and point-dipole approximations \cite{takacs2024accurate}. By using lattice site constraints, we address two main issues. First, in high-dimensional spaces, random walks are inefficient, so by mapping the possible values of the parameters of interest to a discrete set, we can sample from such a discrete set via a random walk rather than allowing the unknown parameters to take on any value in a specified closed domain. Second, because lattice position determines the strength of the hyperfine coupling, and there are lattice sites that, by symmetry, have identical hyperfine couplings, we have further constraints on the number of allowable nuclear spins with identical hyperfine couplings, provided that two walkers are not allowed to simultaneously occupy the same lattice site.

\section{Results}
\label{sec:results}

\subsection{Results from simulations}
Experimentally, the coherence signals are obtained by applying a series of $N$ Carr-Purcell (CP) pulses separated by varying inter-pulse spacings $\tau_j$. Hence, to obtain each additional data point at a new $\tau_j$, the CP experiment must be repeated many times and the results averaged to mitigate noise from photon detection, and each experiment for each data point takes $\propto 2\tau_jN$. We assume we are working with data in a noise and sparsity regime where the inverse problem is ill-posed, and we seek to understand how sparsity and noise impact the accuracy of the recovery procedure by studying the accuracy on simulated data.

\subsection{Computational details}
\label{subsec:comp_details}

For each simulation, we generate coherence signals based on a set of simulated $^{13}$C that determine the number of spins and the hyperfine couplings for specified experimental parameters and relative decoherence values. We hybridize the algorithms discussed in Sec. \ref{sec:algorithms} as outlined in Alg. \ref{alg:hybrid_alg}, and we initialize 5 ensembles and let each ensemble evolve for 25,000 steps, and we consider the first 10,000 steps to be burn-in time. We cycle between 50 steps of RJMCMC to determine the number of spins in the system, 100 steps of PT to obtain a better fit of the hyperfine coupling Hamiltonian parameters for a fixed number of spins, and 25 steps of RWMH for the relative decoherence parameter $\lambda$. For more details and practical strategies to choose and tune hyperparameters, see Appendix \ref{app:hyperparams}.

For the lattice sites and sets of hyperfine parameters computed at each lattice site, we use data from \cite{takacs2024accurate} and define a cutoff of lattice sites by considering only lattice sites with hyperfine couplings with either component of $A_{\parallel}$ or $A_{\perp}$ greater than 5 kHz. 
For RJMCMC, we consider a maximum of $k_{\text{max}} = 50$ nuclear spins to be recovered. For PT, we use $J=10$ strands and inverse temperatures $\beta_j = 2^{-j}$. The discrete random walk over lattice sites uses a walker radius of $R = 5$ \AA, and the continuous RWMH for $D$ over $[0, 1]$ with reflected boundary conditions uses a walker radius of $R = 0.01$. 
Unless stated otherwise, the default values for the simulated experiments are 250 uniformly sampled $\tau_j$ values between 0 and 8 $\mu$s, $N=16$ dynamical decoupling pulses, an external applied magnetic field of 311 G, and $\epsilon \sim \mathcal{N}(0, 0.001)$ noise added to each data point of the simulated coherence signal. 
For details on how hyperparameters are chosen and how they impact the performance of the hybrid algorithm, see Appendix \ref{app:hyperparams}.

We measure the recovery rate of hyperfine parameters by examining the fraction of simulated spins that appeared in the posterior distribution across the entire spin configuration space. A recovery rate of 1 indicates that a given spin appears in every spin configuration, while a rate of 0 signifies that the spin does not appear in any posterior configuration, and we properly account for duplicate and triplicate hyperfine couplings. In the simulations, where both lattice positions and hyperfine couplings are known, we impose a maximum allowed mismatch of $|\Delta A_{\parallel}|, |\Delta A_{\perp}| \leq 0.1$ kHz for a nuclear spin to be detected. This tolerance is not intended to relax the hyperfine constraint, but rather to account for small numerical errors in the DFT-computed hyperfine couplings between symmetry-related lattice sites. More precisely, let $B^{(j)}$ denote the $j$-th posterior spin-bath sample, with $j = 1, \dots, M$, and define an indicator function $\mathbb{I}_i^{(j)}$ for the presence of a given simulated (or experimentally-confirmed) spin $s_i \in S$ posterior sample $B^{(j)}$, where $\mathbb{I}_i^{(j)}$ is 1 if $s_i \in B^{(j)}$ and 0 otherwise. Here, spin identity is determined by matching hyperfine parameters within the mismatch tolerance. The detection rate for an individual spin $s_i$ is defined as
\begin{equation}
    R_i = \frac{1}{M}\sum_{j=1}^{M}\mathbb{I}_{i}^{(j)},
\end{equation}
and the overall detection rate for a given simulated (or experimentally-confirmed) spin-bath $S$ of size $n$ is
\begin{equation}
    R = \frac{1}{n}\sum_{i=1}^n R_i.
\end{equation}
We then define the overall recovery rate by averaging these individual rates across all spins in the bath. When multiple simulated baths are considered (e.g., with different bath sizes), the reported values are further averaged across all simulated baths.

Additionally, we assess the accuracy of recovering the number of spins by comparing the modes of the posterior distribution with the number of simulated spins in each hyperfine coupling grouping. We report the discrepancy in the number of selected spins as |(mode of posterior distribution of $k) - (\#$ of simulated spins)|, so a discrepancy of 0 indicates that the mode of the posterior distribution matches the number of spins that are simulated, and a discrepancy greater than 0 indicates that the Hamiltonian model dimension produced by the hybrid MCMC procedure is different from the number of simulated spins. We quantify false positive rate by considering the modal spin distribution corresponding to the modal model dimension of the posterior distribution. Let $S^*$ denote the set of spins in the modal spin-bath configuration of size $n$, and let $B^{(j)}$ denote the $j$-th posterior spin-bath, with $j = 1, \dots, M$. We define a sample-averaged false absence rate as 
\begin{equation}
    FP = \frac{1}{nM} \sum_{s \in S^*}\sum_{j=1}^{M}\mathbb{I}_{s \notin B^{(j)}}
\end{equation}
where $\mathbb{I}_{s \notin B^{(j)}}$ is a non-occurrence indicator function for the presence of a spin $s$ in nuclear spin bath $B^{(j)}$. This sampled-averaged false absence rate measures the fraction of posterior samples in which spins belonging to the modal configuration are absent. We note that this is a sample-averaged measure rather than a strict false-positive rate in the classical classification sense. All results are grouped by hyperfine coupling magnitude, which we define as $\sqrt{A_{\parallel}^2 + A_{\perp}^2}$, and we average across 16 different simulated spin-bath configurations containing between 5 and 20 simulated spins.

\begin{algorithm}[t]
    \caption{Hybridized MCMC algorithm for recovering local nuclear spin environment}
    \label{alg:hybrid_alg}
    \begin{algorithmic}[1]
        \Statex \textbf{Input}: Data $\mathbf{d}_d$, total number of steps $N_{\text{tot}}$, number of RWMH steps $N_{\text{RWMH}}$, number of RJMCMC steps $N_{\text{RJMCMC}}$, number of PT steps $N_{\text{PT}}$, discrete parameter corresponding to nuclear spins $\mathbf{a}_k = \{p_{d_1}, \dots, p_{d_k}, k\}$, continuous parameters corresponding to relative decoherence $\mathbf{b} = \{ \lambda\}$, fixed parameters in the forward model $\mathbf{c} = \{ N, \tau, B_z\}$, family of Hamiltonians $\{\mathcal{H}_k \}_{k=1}^{k_{\text{max}}}$, likelihood model $\mathcal{L}(\mathbf{d}_d|f_k(\mathbf{a_k}, \mathbf{b}, \mathbf{c}))$, and hyperparameters for the individual MCMC algorithms
        \For{$n = 0, \dots N_{\text{tot}}-1$}
            \State $n_{\text{RWMH}}  \leftarrow 0$
            \State $n_{\text{RJMCMC}} \leftarrow 0 $
            \State $n_{\text{PT}} \leftarrow 0$
            \For{$n_{\text{RWMH}} = 0, \dots, N_{\text{RWMH}}-1$} 
                \State RWMH (Alg. \ref{alg:rwmh}) with 
                \Statex \textbf{Input}:  $\mathbf{p} = \{\lambda\}$, $\mathbf{q} = \{\mathbf{a}_k, k\}, \mathbf{p}^{(0)}=\{\lambda^{n+n_{\text{RWMH}}-1}\}, \mathbf{q}^{(0)}=\{\mathbf{a}_k^{n+n_{\text{RWMH}}-1}, k^{n+n_{\text{RWMH}}-1}\}$ 
                \Statex \textbf{Output}: $\{\mathbf{p}^n\}_{n=n}^{n+N_{\text{RWMH}}}=\{ \lambda^n\}_{n=n}^{n+N_{\text{RWMH}}}$, $\{\mathbf{q}^n\}_{n=n}^{n+N_{\text{RWMH}}}=\{ \mathbf{a}_k^n, k^n\}_{n=n}^{n+N_{\text{RWMH}}}$
                \State $n  \leftarrow n + n_{\text{RWMH}} $
            \EndFor
            \For{$n_{\text{RJMCMC}} = 0, \dots, N_{\text{RJMCMC}}-1$}
                \State RJMCMC (Alg. \ref{alg:rjmcmc}) with 
                \Statex \textbf{Input}:  $\mathbf{p} = \{\mathbf{a}_k\}$, $\mathbf{q} = \{\lambda\}, \mathbf{p}^{(0)}=\{\mathbf{a}_k^{n+n_{\text{RJMCMC}}-1}\}, \mathbf{q}^{(0)}=\{\lambda^{n+n_{\text{RJMCMC}}-1}\}$ 
                \Statex \textbf{Output}: $\{\mathbf{p}^n\}_{n=n}^{n+N_{\text{RJMCMC}}}=\{ \mathbf{a}_k^n\}_{n=n}^{n+N_{\text{RJMCMC}}}$, $\{\mathbf{q}^n\}_{n=n}^{n+N_{\text{RJMCMC}}}=\{ \lambda^n\}_{n=n}^{n+N_{\text{RJMCMC}}}$, $\{k^n\}_{n=n}^{n+N_{\text{RJMCMC}}}$
                \State $n  \leftarrow n + n_{\text{RJMCMC}} $
            \EndFor
            \For{$n_{\text{PT}} = 0, \dots, N_{\text{PT}}-1$}
                \State PT (Alg. \ref{alg:parallel_temp}) with 
                \Statex \textbf{Input}:  $\mathbf{p} = \{\mathbf{a}_k\}$, $\mathbf{q} = \{\lambda, k\}, \mathbf{p}^{(0)}=\{\mathbf{a}_k^{n+n_{\text{PT}}-1}\}, \mathbf{q}^{(0)}=\{\lambda^{n+n_{\text{PT}}-1}, k^{n+n_{\text{PT}}-1}\}$ 
                \Statex \textbf{Output}: $\{\mathbf{p}^n\}_{n=n}^{n+N_{\text{PT}}}=\{ \mathbf{a}_k^n\}_{n=n}^{n+N_{\text{PT}}}$, $\{\mathbf{q}^n\}_{n=n}^{n+N_{\text{PT}}}=\{ \lambda^n, k^n\}_{n=n}^{n+N_{\text{PT}}}$
                \State $n  \leftarrow n + n_{\text{PT}} $
            \EndFor
        \EndFor
        \Statex \textbf{Output}: Append all $\mathbf{a}_k$, $\lambda$, and $k$ to obtain $\{ \mathbf{a}^n_k\}_{n=0}^{N_{\text{tot}}}$, $\{ \lambda^n\}_{n=0}^{N_{\text{tot}}}$, and $\{ k^n\}_{n=0}^{N_{\text{tot}}}$
    \end{algorithmic}
\end{algorithm}

\subsubsection{Performance in sparse data limit}
\label{subsec:sparse_data}

\begin{figure}[t!]
    \centering
    \includegraphics[width=\textwidth]{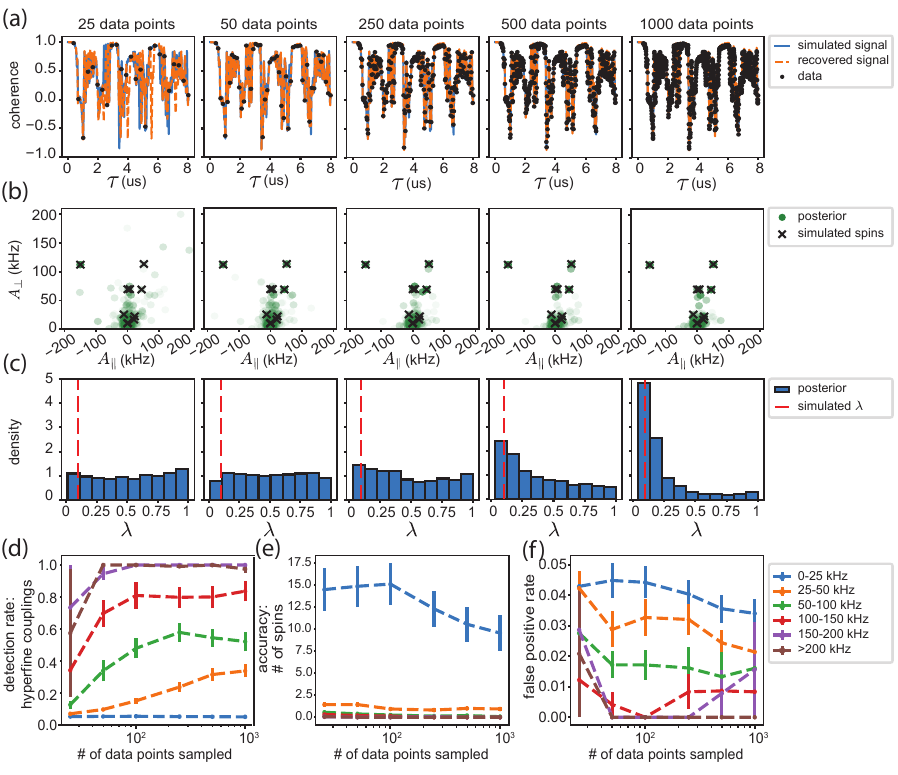} 
    \caption{\textbf{Recovery in sparse data limit}. \textbf{(a)} We simulate a system of ten $^{13}$C surrounding a nitrogen vacancy (NV) center in diamond, and plot the resulting coherence signal for a 16-pulse dynamical decoupling experiment in an external magnetic field of 311 G. We sample a varying number of data points with noise $\epsilon \sim \mathcal{N}(0, 0.001)$, and we plot the coherence signal from the nuclear spin parameters recovered from the specified number of sampled data points. Posterior distribution of the hyperfine components of the nuclear spins \textbf{(b)} and relative decoherence parameter ($\lambda$) \textbf{(c)} recovered from the coherence signal with varying number of data points sampled corresponding to the coherence signal in $\textbf{(a)}$. In \textbf{(d)}, \textbf{(e)}, and \textbf{(f)}, we plot the average results for detection rate of hyperfine couplings, the discrepancy the number of recovered spins in the bath with the number of spins that were simulated, and false positive rate for 16 simulated nuclear spin configurations containing between 5-20 nuclear spins, with the rates plotted by hyperfine magnitude ($\sqrt{A_{\perp}^2 + A_{\parallel}^2}$) of the spins.}
    \label{fig:time_density}
\end{figure}

To evaluate the performance of our recovery method in the limit of sparse data, we vary the number of $\tau_j$ sampled from the coherence signals from a simulated spin bath of ten nuclear spins (Fig. \ref{fig:time_density}a). We then assess whether the posterior distributions of the spin configurations (Fig. \ref{fig:time_density}b) and relative decoherence (Fig. \ref{fig:time_density}c) obtained from the hybrid algorithm described in Sec. \ref{subsec:hybrid_alg} successfully capture the simulated nuclear spin configurations and relative decoherence parameter $\lambda$. To ensure these results generalize across spin configurations, we analyze the recovery rate for the hyperfine couplings, the detection rate for the number of spins, and the false positive rate for nuclear detection for 16 different simulated spin baths containing 5 to 20 spins, with a hyperfine magnitude greater than 5 kHz (Fig. \ref{fig:time_density}d). 

For the sample bath consisting of 10 nuclear spins and varying number of $\tau_j$, we observe, as expected, that the fidelity of the recovered signal improves with an increased number of sampled $\tau_j$ (Fig. \ref{fig:time_density}a). However, we are not necessarily interested in reconstructing the coherence signal, but rather we seek to recover the underlying nuclear spin bath configuration that gives rise to the observed coherence signal. We find that even for signals sampled as sparsely as 50 $\tau_j$, the posterior distribution of hyperfine couplings of spins contains all ten of the simulated spins (Fig. \ref{fig:time_density}b), and the signal sampled as sparsely as 25 $\tau_j$ recovers the nuclear spins with the strongest hyperfine magnitudes. As expected, the more sparsely sampled signals result in more fictitious spins in the posterior compared to more densely sampled spins. The density of the time-point sampling is related to the ill-posedness of the inverse problem; from this sample bath, we also observe that the ability to resolve hyperfine couplings from sparse signals is highly correlated with the strength of the hyperfine magnitude. That is, the strongly coupled spins are clearly resolved in the posterior distributions in Fig. \ref{fig:time_density}b, whereas we are unable to clearly resolve the more weakly coupled spins, even at relatively dense samplings of the signals. We emphasize that the failure to resolve weakly coupled spins is related to the inherence ill-posedness of the inverse problems. Weakly coupled spins lead to smaller dips in the fringes of the coherence signal; as a result, many different combinations of weakly coupled spins can give rise to identical patterns of peaks in the coherence signal, making it impossible to resolve individual weakly coupled spins from a single sparse coherence signal originating from a dynamical decoupling experiment.

While we are able to recover the stronger nuclear spin hyperfine couplings, even from sparse time-point samplings, we observe that relatively higher resolutions of data are required to recover the relative decoherence parameter, $\lambda$, as seen in Fig. \ref{fig:time_density}c. For samplings below 250 $\tau$, we obtain a relatively uniform posterior distribution of $\lambda$, whereas for simulated data with more than 250 $\tau$, the recovered value of $\lambda$ (that is, the mode of the posterior distribution) matches the simulated value. Since $\lambda$ is a decay parameter, extremal peak values of the coherence signal are required to capture the decay accurately. For sparsely sampled data, there is a lower probability that extremal peak values will be sampled, resulting in a relatively uniform distribution of relative decoherence values (Fig. \ref{fig:time_density}c). In contrast, dense samplings of the data will result in many extremal peak values, resulting in a non-uniform posterior distribution of relative decoherence values centered on the simulated value.

To generalize the results beyond a single spin bath, we look at average values of detection rate (Fig. \ref{fig:time_density}d), discrepancy in the number of recovered spins (Fig. \ref{fig:time_density}e), and false positive rates (Fig. \ref{fig:time_density}f). We find that the recovery rate of hyperfine couplings improves with increasing $\tau$, but plateaus beyond approximately 100 sampled time points. Additional sampling beyond this threshold does not significantly enhance the recovered information, indicating a limit on the impact of increasing the rate of sampling in terms of improving the recoverable information content from specific experimental parameters. That is, beyond around 100 sampled $\tau_j$, the ability to recover hyperfine couplings from this coherence signal is limited not by the data resolution but by the experimental values itself, such as the number of CP pulses $N$, the external magnetic field $B_z$, or the maximum $\tau_j$ sampled. When considering the discrepancy in the number of spins, we observe a similar trend: recovery rates increase with the number of sampled time points, eventually saturating beyond 500 time points.

The accuracy of recovery rates is notably lower for weakly coupled nuclei ($<25$ kHz) across multiple metrics, including detection rate, number of recovered spins, and false positive rate, as seen in Fig. \ref{fig:time_density}d-f. This behavior arises from the inherent ill-posedness of the problem in the sparse data regime: distinct configurations of weakly coupled spins can generate identical coherence signals with comparable likelihoods, limiting the ability to resolve individual weakly-coupled nuclear spins. This limitation is a fundamental consequence of the ill-posedness of this inverse problem in the sparse data regime rather than a failure of the hybrid algorithm to accurately resolve the simulated nuclear spin configuration.

In contrast, for nuclear spins with hyperfine couplings exceeding 150 kHz, the hybrid MCMC recovery method exhibits consistently high accuracy. For spins with intermediate coupling strengths, accuracy improves with increased sampling, particularly in the range of 25 to 100 sampled time points (Fig. \ref{fig:time_density}d-f). However, beyond 100 sampled points, the performance of the hybrid algorithm is stable, with detection rates, recovered spin counts, and false positive rates remaining largely unchanged up to 1000 sampled points. This hybrid recovery algorithm is robust even when applied to coherence signals sampled as sparsely as 100 time points.

\subsubsection{Performance in noisy data limit}
\label{subsec:noisy_data}
\begin{figure}[t!]
    \centering
    \includegraphics[width=\textwidth]{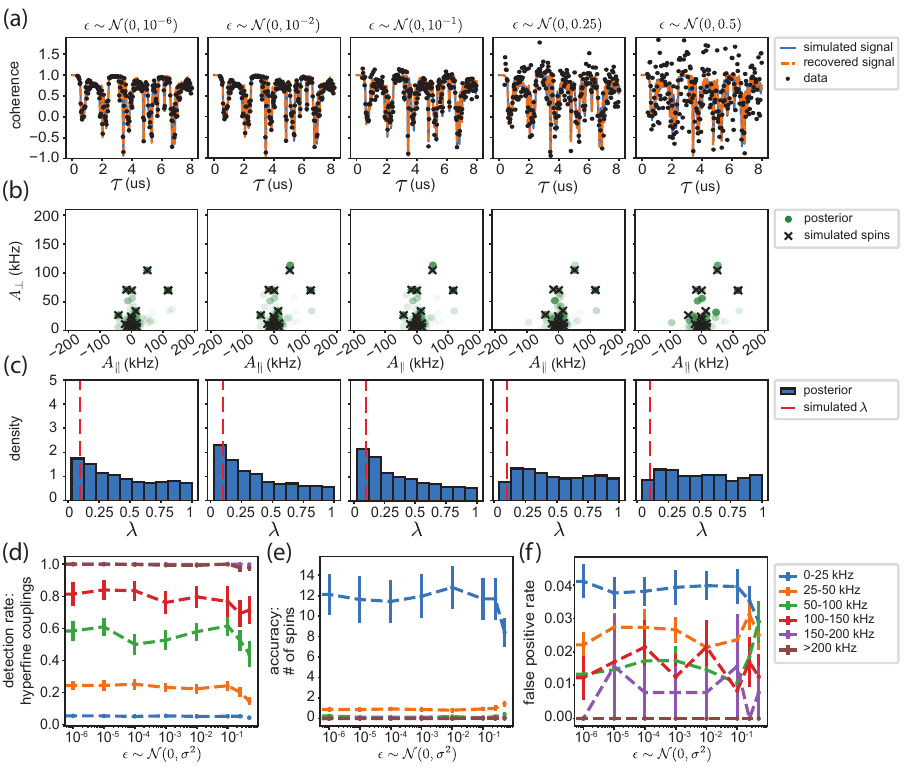} 
    \caption{\textbf{Recovery in noisy data limit}. \textbf{(a)} We simulate a system of seventeen $^{13}$C surrounding a nitrogen vacancy (NV) center in diamond, and plot the resulting coherence signal sampled uniformly for 250 $\tau_j$ for a 16-pulse dynamical decoupling experiment in an external magnetic field of 311 G. We add varying amounts of noise to each data point, and we plot the coherence signal from the nuclear spin parameters recovered from the specified number of sampled data points. We plot the posterior distribution of the hyperfine components of the nuclear spins \textbf{(b)} and relative decoherence parameter ($\lambda$) \textbf{(c)} recovered from the coherence signal with varying number of data points sampled corresponding to the coherence signal in $\textbf{(a)}$. In \textbf{(d)}, \textbf{(e)}, and \textbf{(f)}, we plot the average results for detection rate of hyperfine couplings, the discrepancy the number of recovered spins in the bath with the number of spins that were simulated, and false positive rate for 16 simulated nuclear spin configurations containing between 5-20 nuclear spins, with the rates plotted by hyperfine magnitude ($\sqrt{A_{\perp}^2 + A_{\parallel}^2}$) of the spins.}
    \label{fig:noise_simulation}
\end{figure}

Since experimentally, coherence signals are measured by averaging photon counts across repeats of a single given CP experimental pulse, the shot noise per data point can be decreased by a quadratically longer time average (i.e.,  reducing shot noise by half requires repeating an experiment four times as many repeats). Since averaging takes time, we are interested in how the accuracy of the recover procedure is impacted by the amount of noise. Remarkably, our approach demonstrated high robustness to shot noise across various magnitudes of hyperfine couplings. In an example simulated bath containing seventeen nuclear spins with varying amounts of noise added to each data point, we observe that even for significant amounts of noise ($\epsilon \sim \mathcal{N}(0, 0.25)$), we are able to recover both the coherence signal (Fig. \ref{fig:noise_simulation}a) and the strongly coupled nuclear spins are well-resolved in the posterior distribution of the recovered hyperfine couplings (Fig. \ref{fig:noise_simulation}b). Recovery performance only showed significant degradation when the average shot noise per data point exceeded 0.25, which is substantial given that the coherence signal is bounded between -1 and 1. Even with average noise per data point up to 0.1 , the algorithm was able to recover hyperfine couplings and spin numbers, especially for spins with stronger hyperfine couplings. Furthermore, we find that, similar to decreasing the number of sampled time points, adding an increasing amount of noise results in a more uniform distribution of the posterior distribution of the relative decoherence parameter, $\lambda$ (Fig. \ref{fig:noise_simulation}c). Since the relative decoherence relies on accurate values of the coherence at its peaks, adding an increasing amount of noise results in inaccurate peak values, leading to an inability to recover $\lambda$.  

Over a range of simulated nuclear spin baths, we find that this hybrid recovery approach is remarkably stable to noise added to the simulated coherence data up to $\epsilon \sim \mathcal{N}(0, 0.1)$ in terms of multiple metrics, including the detection rate of the hyperfine couplings (Fig. \ref{fig:noise_simulation}d), discrepancy in number of spins recovered (Fig. \ref{fig:noise_simulation}e), and false positive rate (Fig. \ref{fig:noise_simulation}f). These results suggest that our method is well-suited for noisy data and does not require a separate de-noising step, which is a common requirement in machine learning-based approaches \cite{jung2021deep}. Furthermore, the algorithm's ability to recover nuclear spins across different magnitudes for average shot noise per data point ranging from $10^{-6}$ to $10^{-1}$ indicates that the ability to recover nuclear spins is limited by the information inherent in short, sparsely sampled dynamical decoupling experiments, rather than by shot noise obscuring the hyperfine coupling information present in the coherence signal.

\subsection{Validation on experimental data}
\label{subsec:validation}
\begin{figure}[t!]
    \centering
    \includegraphics[width=\textwidth]{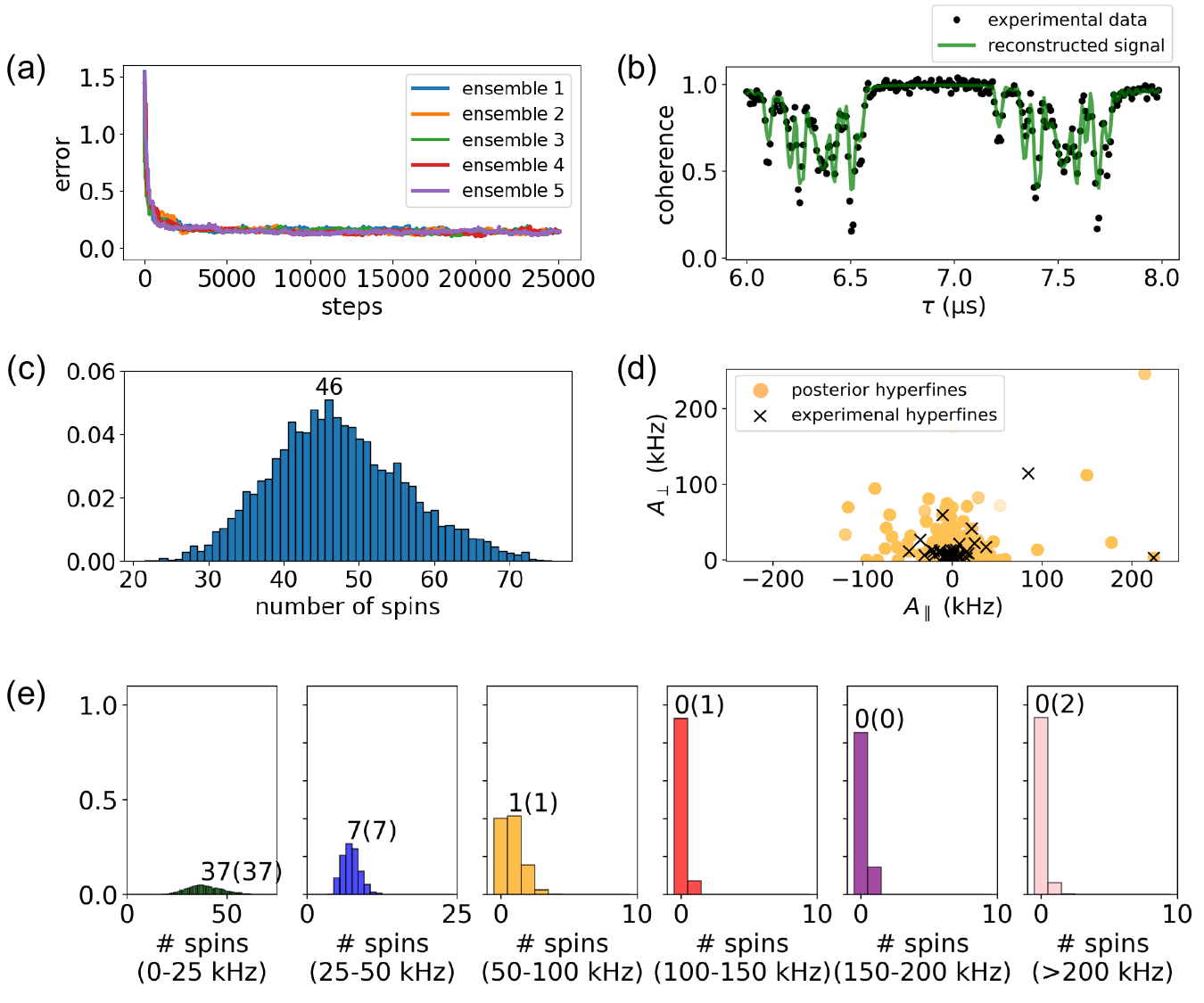} 
   \caption{
\textbf{Validation of hybrid MCMC approach on experimental data:} 
(a) Average error per data point as a function of the number of MCMC steps, with five random ensembles initialized for 25,000 steps each; the first 10,000 steps are treated as burn-in, and the remaining 15,000 steps are the posterior spin configurations. 
(b) Experimental coherence data (250 data points from 6~$\mu$s to 8~$\mu$s of a 32-pulse CPMG sequence) to which we applied the hybrid MCMC algorithm, and the reconstructed signal using the best-fit spin configuration from the posterior distribution. 
(c) Posterior distribution of the total number of spins in the recovered spin configurations. 
(d) Posterior distribution of hyperfine couplings, with the 50 reported nuclear spin hyperfine couplings from the reference study overlaid. 
(e) Posterior distribution of the number of spins by spin magnitude (\(\sqrt{A_\perp^2 + A_\parallel^2}\)), with the mode displayed and the reference value reported in parentheses.
}
    \label{fig:validation}
\end{figure}

We validate the hybrid MCMC approach by applying it to coherence data collected from a single NV center at a temperature of 3.7 K using a 32-pulse CPMG pulse sequence, with $\tau$ ranging from 6 to 8 $\mu$s. For more details on experimental procedure, see App. \ref{app:exp_procedure}. This NV center has well-characterized hyperfine interactions and lattice positions of surrounding nuclear spins, determined from high-resolution multidimensional spectroscopy and correlated sensing experiments \cite{abobeih2019atomic, van2024mapping}. These experimentally determined values serve as reference values.

We use the hyperfine couplings and lattice positions calculated in \cite{takacs2024accurate} as theoretical reference values. The alignment procedure between these theoretical values and the experimentally determined values is described in Appendix \ref{app:detect_rates}. The hybrid MCMC algorithm (Alg. \ref{alg:hybrid_alg}) is initialized with five independent ensembles, each comprising 25,000 MCMC steps. All ensembles converged to the same error regime within 5,000 steps (Fig. \ref{fig:validation}a), and the initial 10,000 steps were discarded as burn-in. The posterior distribution was constructed from the remaining 15,000 steps.

To incorporate additional geometric information from the experimental coherence decay data, we modified the likelihood function from a standard squared-error form (Eq. \ref{eq:l2_likelihood}) to a weighted likelihood function that balances point-wise residuals with global shape differences in the signal. Specifically, we introduce a regularized form that incorporates the squared 2-Wasserstein distance between the normalized empirical data distribution and the predicted signal distribution:
\begin{align}
\mathcal{L}_{\text{mod}}(\mathbf{d}_d | \theta) = 
(1 - \zeta) \cdot \exp\left(
-\frac{1}{2 \sigma^2} \sum_{i=1}^d \left(\mathbf{d}_i - f_k(\mathcal{H}_k(\theta), \tau_i)\right)^2 
\right)
- \zeta \cdot W_2^2\left(f_k(\mathcal{H}_k(\theta)), \mathbf{d}\right)
\end{align}
where $W_2^2$ denotes the squared 2-Wasserstein distance between the empirical data distribution $\mathbf{d}$ and the predicted model distribution $f_k(\mathcal{H}_k(\theta))$, and $\zeta$ is a weighting parameter.

This hybrid formulation enables the inference algorithm to capture both local pointwise agreement and global distributional similarity. The squared-error term ensures that the model fits the observed decay values at each delay time $\tau_i$, while the Wasserstein term penalizes global mismatches in the overall structure of the coherence decay curve—an important feature when dealing with noisy, underdetermined, or aliased experimental data \cite{engquist2016optimal}. This additional regularization is not required when testing our hybrid MCMC approach on simulated data generated from known hyperfine parameters. In the simulation case, the model is sampled from the same distribution as the simulated data, and there is no error due to mismatches between the model and physical reality. Therefore, a standard likelihood based on squared residuals is sufficient for an accurate reconstruction, in the case of the simulated data. In contrast, when analyzing experimental data, we know that the true hyperfine couplings differ systematically from the theoretical values used to define the sampling space \cite{takacs2024accurate}. As a result, directly comparing the predicted and experimental signals point-wise can lead to overfitting or biased inference. Incorporating the Wasserstein distance allows the inference process to remain robust to such mismatches by focusing on the agreement in the shape and distribution of the coherence signal, rather than on the exact point-wise correspondence. The weight $\zeta = 0.5$ was selected after benchmarking several values, as detailed in Appendix~\ref{app:validate_zeta}, where we show that this value of $\zeta$ balances local fit and global structure matching.

The reconstructed coherence signal (Fig. \ref{fig:validation}b) captures many of the experimentally observed features, particularly the smaller peaks, but underestimates the depths of the strongest peaks. This is consistent with the known limitations in the accuracy of theoretical hyperfine calculations, particularly for strongly coupled nuclear spins. For more details and a discussion on the accuracy dependence of the hybrid MCMC approach on the accuracy of the input hyperfine couplings computed with DFT, see App. \ref{app:dft_hfs}.

The posterior distribution of the total number of nuclear spins (Fig. \ref{fig:validation}c), with a modal value of 46, shows excellent agreement with the experimental reference value of 50. We further evaluated the posterior distribution of hyperfine couplings (Fig. \ref{fig:validation}d). A total of 45 out of the 48 experimentally identified nuclear spins are detected in the posterior distribution, with 22 of these spins exhibiting detection rates exceeding their baseline probabilities (Table \ref{tab:hf_compare}). The calculation of detection rates and baseline probabilities is detailed in Appendix \ref{app:detect_rates}. One strongly coupled spin (labeled C36 in Table 2 with ($A_{\parallel}, A_{\perp}$) = (84.52, 114.42) kHz) is not detected in the posterior distribution (see Fig. \ref{fig:validation}d). This illustrates the combined effects of sparse experimental data and known discrepancies between DFT-computed and experimentally measured hyperfine couplings: when the theoretical hyperfine values deviate from experiment, the posterior detection probability can be reduced even for otherwise well-resolved spins. Appendix \ref{app:dft_hfs} provides a detailed analysis showing that recovery rates for both spin number and hyperfine couplings remain high when the input couplings are accurate to within $\sim 1$ kHz, while larger discrepancies between the DFT-computed and experimentally-measured hyperfine couplings, as in the case of C36, lead to decreased detection.

Our approach offers significant advantages over prior methods in terms of data efficiency, computational cost, and adaptability. A machine learning method proposed in Ref. \cite{jung2021deep} identified 23 nuclear spins in the same experimental dataset but required ~14 hours of computation using both CPU and GPU resources (an Intel Core i9-9920X with 12 cores and 2× NVIDIA RTX 2080 Ti), along with 11,000 data points for a 32-pulse experiment (and 7,500 for a 256-pulse one). Another ML-based method for this same inverse problem \cite{varona2024automatic}, which to the best of our knowledge has not yet been experimentally validated, uses 3.6 million synthetic training samples, each of which requires a forward model evaluation of the spin coherence model. Although they do not report computational time, training data generation alone involves significantly more model evaluations than our hybrid approach, which requires approximately 750,000 forward evaluations in total (25,000 MCMC walker steps per ensemble with an additional approximately 125,000 evaluations per ensemble during parallel tempering, with 5 ensembles run in total). Furthermore, their model must be retrained when experimental conditions, such as magnetic field or pulse sequences, are changed.

In contrast, our method achieves comparable performance using only 250 data points from a single 32-pulse experiment and runs in approximately 8 hours using only 5-cores of a single CPU node (Intel Xeon E5-2680 v4), without requiring pre-training or additional tuning when experimental conditions are changed. The experimental dataset used here was originally acquired for different studies \cite{abobeih2019atomic, van2024mapping}; we sub-sampled the available time points to test the inference framework in a regime currently inaccessible to prior inverse-problem techniques, demonstrating that accurate spin detection can be achieved with substantially fewer measurements. Typical acquisition of a single 32-pulse coherence signal with 250 time points requires roughly 5–10 hours under standard laboratory conditions, including NV selection, alignment, and calibration, making the measurement time comparable to the runtime of the inference algorithm. Compared to another non-ML approach proposed in Ref. \cite{oh2020algorithmic}, which identified 14 experimental spins using high-resolution data (around 10,000 time points of the coherence signal) across multiple $^{13}$C Larmor periods, our method identifies more spins using a much sparser dataset. The tradeoff of working with sparse data is the persistence of ill-posedness; however, our Bayesian inference framework reduces the number of plausible lattice sites from 3,518 to 230 by leveraging posterior detection probabilities.

Finally, the hybrid MCMC method accurately characterizes the distribution of nuclear spins across different hyperfine magnitude regimes (Fig. \ref{fig:validation}e). Specifically, the modes of the posterior distributions for weakly coupled spins (<100 kHz) align exactly with reference values, while the number of strongly coupled spins is slightly underestimated. This outcome is consistent with the known limitations of the theoretical hyperfine values (see Appendix \ref{app:dft_hfs}).

\section{Discussion}
\label{sec:discussion}

Hybridized Markov Chain Monte Carlo (MCMC) methods offer a flexible and powerful framework for addressing trans-dimensional inverse problems, particularly under ill-posed conditions. Although such hybrid approaches, where different MCMC strategies are combined to enhance sampling efficiency, have found broad utility in fields such as ecology \cite{walker2006stochastic}, evolutionary biology \cite{mathew2012bayesian}, and geophysics \cite{zhang2021geological, reuschen2021efficient}, their application to quantum systems presents unique advantages. This is largely due to the discrete nature of quantum states and the hierarchical structure of prior knowledge that often accompanies physical parameters, whether derived from experimental measurements or from first-principles calculations. When prior knowledge strongly constrains the possible values of certain parameters, such as nuclear spin quantum numbers or symmetry-allowed configurations, it is advantageous to represent these parameters as discrete, thereby restricting the search space to physically admissible values and reducing computational cost. Conversely, for parameters where the uncertainty is significant, a continuous representation may be more appropriate, allowing the model to explore a broader and less constrained space informed by physically motivated probability distributions. This dual treatment of model parameters—as either discrete or continuous depending on the level of prior information—provides a systematic means of integrating \textit{a priori} knowledge into the sampling process, balancing computational efficiency with physical rigor.

Crucially, the hybrid MCMC framework does not depend on the availability of DFT-derived hyperfine tensors and remains fully operational even in their absence. When the crystal lattice is known but detailed electronic-structure information is unavailable, the same hybrid sampling strategy can be applied over discrete lattice sites using magnetic dipolar hyperfine interactions computed analytically and directly from the fixed lattice geometry. This lattice-constrained formulation captures the dominant distance- and orientation-dependent physics governing the nuclear–electron interaction while substantially reducing the dimensionality of the inference problem. A continuous RWMH scheme can then relax the hyperfine coupling parameters associated with each lattice site around their approximate dipolar values, without relaxing the underlying lattice positions themselves. This hybrid inference strategy establishes the approach as a broadly applicable and practical technique for reconstructing nuclear spin environments across a wide class of spin-defect systems, including semiconductors for which reliable DFT hyperfine calculations are unavailable or impractical to compute. More generally, in systems where even lattice-resolved prior information is unavailable, the same Bayesian framework can be implemented in a fully continuous trans-dimensional setting using a continuous formulation of RJMCMC \cite{green2009reversible} or continuous RWMH sampling over both the number and spatial coordinates of nuclear spins without requiring any discrete lattice constraints.

A central challenge in applying Markov Chain Monte Carlo (MCMC) methods to quantum systems lies in the high computational cost associated with forward models, which map parameters of interest to observable quantities. In quantum materials and molecular systems, such forward models are often based on \textit{ab initio} electronic structure methods, such as density functional theory (DFT), whose computational cost typically scales as $O(N^3)$ with system size $N$. For certain physical properties that require even more computationally expensive methods, such as coupled-cluster or multireference methods, the scaling becomes substantially worse—up to $O(N^6)$ or beyond—rendering them impractical for large systems or for problems requiring extensive sampling. In trans-dimensional Bayesian inference, MCMC algorithms often require tens of thousands of likelihood evaluations to adequately sample posterior distributions. If each of these evaluations involves a costly first principles calculation, the cumulative computational burden becomes prohibitive. Thus, the development of surrogate models, reduced-order methods, or data-driven approximations that retain quantum mechanical fidelity while significantly lowering computational overhead is essential for enabling scalable inference of the properties of quantum systems.

One promising strategy to address the computational bottleneck is the use of machine learning (ML) to construct surrogate models for forward simulations. Rather than attempting to learn the inverse map from observables to model parameters—which is fundamentally ill-posed and highly sensitive to noise—ML approaches can be employed to approximate the forward model itself, which is well-posed and grounded in the underlying physics. Surrogate models trained on a limited set of high-fidelity quantum mechanical calculations can reproduce system observables with sufficient accuracy while drastically reducing the computational cost of individual likelihood evaluations. This approach enables efficient sampling within MCMC frameworks, without compromising the interpretability or uncertainty quantification inherent to Bayesian inference. Recent studies have demonstrated the effectiveness of surrogate ML models in capturing structure–property relationships in a range of quantum systems, including the prediction of spectroscopic signatures, electronic properties, and energy landscapes \cite{nguyen2024efficient, babbar2024explainability, nyshadham2019machine, verriere2022building, roongcharoen2024machine, chandrasekaran2019solving}. The integration of such models into inference workflows provides a scalable path forward for the characterization of the properties of quantum materials and molecular systems, especially in regimes where direct simulation remains computationally prohibitive.

A notable strength of the Bayesian framework, and hybridized MCMC approaches in particular, lies in their systematic improvability and conceptual flexibility. Enhancements in the accuracy or efficiency of forward models—whether through advances in numerical algorithms or the introduction of machine-learned surrogates—translate directly into improved sampling performance and more accurate inference, without requiring fundamental changes to the underlying methodology. Moreover, physical constraints, conservation laws, and symmetries can be naturally and rigorously incorporated into the likelihood function, allowing the inference process to remain firmly grounded in the physics of the problem. Modifying the forward model or refining the likelihood to reflect improved understanding or additional experimental input incurs minimal overhead, enabling a modular and extensible workflow. This stands in contrast to direct machine learning approaches, which typically require retraining on new datasets and often involve nontrivial engineering of loss functions to enforce physical consistency. In this sense, Bayesian inference provides a transparent and adaptable framework for the characterization of many properties of quantum systems, capable of evolving alongside theoretical and computational advances.

A potential limitation of the hybrid MCMC approach arises when the true Hamiltonian does not lie within the predefined class of candidate models. However, the framework is naturally compatible with established techniques for model selection and averaging, which can mitigate issues related to model misspecification. Bayesian model averaging \cite{clyde2011bayesian}, mixture models \cite{seni2010ensemble}, and ensemble strategies \cite{buhlmann2011bagging} such as bagging offer systematic approaches to incorporate uncertainty over model structure. In particular, bagging methods—originally developed in the context of machine learning—construct ensembles by training models on bootstrapped subsets of the data and combining their predictions through voting, averaging, or regularized aggregation to improve robustness; such procedures are adopted in sparse data regimes where small variations in the dataset may lead to significant shifts in the inferred posterior \cite{adrian2024stabilizing}. By enabling principled treatment of both parameter and model uncertainty, such extensions could preserve the interpretability and physical grounding of the Bayesian approach while enhancing its stability and generalizability. Nonetheless, if the true model lies far outside the chosen candidate set, the current framework cannot generate a fully data-driven model; some choice of forward model or candidate class is always required. This feature distinguishes our approach from purely data-driven or machine learning–based methods. At the same time, leveraging substantial prior information about the physical system, such as Hamiltonian structure and DFT-based priors, is precisely what allows us to extract meaningful information from this highly ill-posed inverse problem.

Another potential limitation of MCMC-based strategies is their computational cost, which can become significant for high-dimensional models. This is an inherent feature of sampling-based approaches: the walkers require a burn-in period to explore the posterior distribution before producing reliable samples, and each likelihood evaluation can be expensive, particularly when forward models involve complex physics or first-principles calculations. While surrogate models \cite{li2023surrogate} and reduced-order approximations \cite{lieberman2010parameter} can mitigate the computational burden, the fundamental requirement for a sufficiently long burn-in and adequate sampling cannot be eliminated. High-dimensional problems require more computational resources, and careful attention must be paid to balancing model complexity, sampling efficiency, and available computational power.

Future developments of hybrid MCMC methods for ill-posed inverse problems will benefit from several complementary advances. Adaptive hybridization schemes that dynamically toggle between discrete and continuous parameter spaces in response to data quality and uncertainty could significantly enhance sampling efficiency and robustness. The integration of surrogate models—potentially guided by active learning protocols—promises not only to accelerate convergence but also to enable real-time Bayesian updating, laying the foundation for autonomous, data-driven experimental design and continuous data assimilation. Moreover, realizing the full potential of these approaches will require efficient, scalable implementations of parallel and distributed MCMC algorithms capable of exploiting modern high-performance computing architectures. Coupled with user-friendly software frameworks that facilitate hybridization across diverse MCMC strategies, such advances will be crucial to the widespread adoption of hybrid MCMC for rapid and scalable quantum characterization across a broad range of physical quantum systems.

\section*{Data and code availability}
Data and supporting code that support this study will be made available through github: \texttt{https://github.com/pabigail/rjmcmc-hyperfine-recovery}.

\section*{Acknowledgments}
We thank F. Joseph Heremans, Mykyta Onizhuk, Daniel Sanz-Alonso, Nathan Waniorek, Christopher Egerstrom, Benjamin Pingault, and Christina Ioannou for useful discussions.
 A.N.P. acknowledges support from the DOE CSGF under Award Number DE-SC0022158. This research used resources of the University of Chicago Research Computing Center. This work was supported by AFOSR Grant No. FA9550-22-1-0370.


\begin{thebibliography}{56}
\providecommand{\natexlab}[1]{#1}
\providecommand{\url}[1]{\texttt{#1}}
\expandafter\ifx\csname urlstyle\endcsname\relax
  \providecommand{\doi}[1]{doi: #1}\else
  \providecommand{\doi}{doi: \begingroup \urlstyle{rm}\Url}\fi

\bibitem[Castelletto and Boretti(2020)]{castelletto2020silicon}
Stefania Castelletto and Alberto Boretti.
\newblock Silicon carbide color centers for quantum applications.
\newblock \emph{Journal of Physics: Photonics}, 2\penalty0 (2):\penalty0
  022001, 2020.
\newblock \doi{10.1088/2515-7647/ab77a2}.

\bibitem[Rodgers et~al.(2021)Rodgers, Hughes, Xie, Maurer, Kolkowitz,
  Bleszynski~Jayich, and de~Leon]{rodgers2021materials}
Lila~VH Rodgers, Lillian~B Hughes, Mouzhe Xie, Peter~C Maurer, Shimon
  Kolkowitz, Ania~C Bleszynski~Jayich, and Nathalie~P de~Leon.
\newblock Materials challenges for quantum technologies based on color centers
  in diamond.
\newblock \emph{MRS Bulletin}, 46\penalty0 (7):\penalty0 623--633, 2021.
\newblock \doi{10.1557/s43577-021-00137-w}.

\bibitem[Anderson et~al.(2022)Anderson, Glen, Zeledon, Bourassa, Jin, Zhu,
  Vorwerk, Crook, Abe, Ul-Hassan, Ohshima, Nguyen, Galli, and
  Awschalom]{anderson2022five}
Christopher~P Anderson, Elena~O Glen, Cyrus Zeledon, Alexandre Bourassa,
  Yu~Jin, Yizhi Zhu, Christian Vorwerk, Alexander~L Crook, Hiroshi Abe, Jawad
  Ul-Hassan, Takeshi Ohshima, T.~Son Nguyen, Giulia Galli, and David~D
  Awschalom.
\newblock Five-second coherence of a single spin with single-shot readout in
  silicon carbide.
\newblock \emph{Science Advances}, 8\penalty0 (5):\penalty0 eabm5912, 2022.
\newblock \doi{10.1126/sciadv.abm5912}.

\bibitem[Kanai et~al.(2022)Kanai, Heremans, Seo, Wolfowicz, Anderson, Sullivan,
  Onizhuk, Galli, Awschalom, and Ohno]{kanai2022generalized}
Shun Kanai, F~Joseph Heremans, Hosung Seo, Gary Wolfowicz, Christopher~P
  Anderson, Sean~E Sullivan, Mykyta Onizhuk, Giulia Galli, David~D Awschalom,
  and Hideo Ohno.
\newblock Generalized scaling of spin qubit coherence in over 12,000 host
  materials.
\newblock \emph{Proceedings of the National Academy of Sciences}, 119\penalty0
  (15):\penalty0 e2121808119, 2022.
\newblock \doi{10.1073/pnas.2121808119}.

\bibitem[Bourassa et~al.(2020)Bourassa, Anderson, Miao, Onizhuk, Ma, Crook,
  Abe, Ul-Hassan, Ohshima, Son, Galli, and Awschalom]{bourassa2020entanglement}
Alexandre Bourassa, Christopher~P Anderson, Kevin~C Miao, Mykyta Onizhuk,
  He~Ma, Alexander~L Crook, Hiroshi Abe, Jawad Ul-Hassan, Takeshi Ohshima,
  Nguyen~T Son, Giulia Galli, and David~D Awschalom.
\newblock Entanglement and control of single nuclear spins in isotopically
  engineered silicon carbide.
\newblock \emph{Nature Materials}, 19\penalty0 (12):\penalty0 1319--1325, 2020.
\newblock \doi{10.1038/s41563-020-00802-6}.

\bibitem[Bradley et~al.(2022)Bradley, de~Bone, M{\"o}ller, Baier, Degen,
  Loenen, Bartling, Markham, Twitchen, Hanson, Elkouss, and
  Taminiau]{bradley2022robust}
CE~Bradley, SW~de~Bone, PFW M{\"o}ller, S~Baier, MJ~Degen, SJH Loenen,
  HP~Bartling, M~Markham, DJ~Twitchen, R~Hanson, D~Elkouss, and TH~Taminiau.
\newblock Robust quantum-network memory based on spin qubits in isotopically
  engineered diamond.
\newblock \emph{npj Quantum Information}, 8\penalty0 (1):\penalty0 122, 2022.
\newblock \doi{10.1038/s41534-022-00637-w}.

\bibitem[Waeber et~al.(2019)Waeber, Gillard, Ragunathan, Hopkinson, Spencer,
  Ritchie, Skolnick, and Chekhovich]{waeber2019pulse}
AM~Waeber, G~Gillard, G~Ragunathan, M~Hopkinson, P~Spencer, DA~Ritchie,
  MS~Skolnick, and EA~Chekhovich.
\newblock Pulse control protocols for preserving coherence in dipolar-coupled
  nuclear spin baths.
\newblock \emph{Nature Communications}, 10\penalty0 (1):\penalty0 3157, 2019.
\newblock \doi{10.1038/s41467-019-11160-6}.

\bibitem[Dong et~al.(2020)Dong, Calderon-Vargas, and Economou]{dong2020precise}
Wenzheng Dong, FA~Calderon-Vargas, and Sophia~E Economou.
\newblock Precise high-fidelity electron--nuclear spin entangling gates in nv
  centers via hybrid dynamical decoupling sequences.
\newblock \emph{New Journal of Physics}, 22\penalty0 (7):\penalty0 073059,
  2020.
\newblock \doi{10.1088/1367-2630/ab9bc0}.

\bibitem[Cramer et~al.(2016)Cramer, Kalb, Rol, Hensen, Blok, Markham, Twitchen,
  Hanson, and Taminiau]{cramer2016repeated}
Julia Cramer, Norbert Kalb, M~Adriaan Rol, Bas Hensen, Machiel~S Blok, Matthew
  Markham, Daniel~J Twitchen, Ronald Hanson, and Tim~H Taminiau.
\newblock Repeated quantum error correction on a continuously encoded qubit by
  real-time feedback.
\newblock \emph{Nature Communications}, 7\penalty0 (1):\penalty0 11526, 2016.
\newblock \doi{10.1038/ncomms11526}.

\bibitem[Taminiau et~al.(2012)Taminiau, Wagenaar, Van~der Sar, Jelezko,
  Dobrovitski, and Hanson]{taminiau2012detection}
TH~Taminiau, JJT Wagenaar, T~Van~der Sar, Fedor Jelezko, Viatcheslav~V
  Dobrovitski, and R~Hanson.
\newblock Detection and control of individual nuclear spins using a weakly
  coupled electron spin.
\newblock \emph{Physical Review Letters}, 109\penalty0 (13):\penalty0 137602,
  2012.
\newblock \doi{10.1103/PhysRevLett.109.137602}.

\bibitem[Taminiau et~al.(2014)Taminiau, Cramer, van~der Sar, Dobrovitski, and
  Hanson]{taminiau2014universal}
Tim~H Taminiau, Julia Cramer, Toeno van~der Sar, Viatcheslav~V Dobrovitski, and
  Ronald Hanson.
\newblock Universal control and error correction in multi-qubit spin registers
  in diamond.
\newblock \emph{Nature Nanotechnology}, 9\penalty0 (3):\penalty0 171--176,
  2014.
\newblock \doi{10.1038/nnano.2014.2}.

\bibitem[Marcks et~al.(2024)Marcks, Onizhuk, Delegan, Wang, Fukami, Watts,
  Clerk, Heremans, Galli, and Awschalom]{marcks2024guiding}
Jonathan~C Marcks, Mykyta Onizhuk, Nazar Delegan, Yu-Xin Wang, Masaya Fukami,
  Maya Watts, Aashish~A Clerk, F~Joseph Heremans, Giulia Galli, and David~D
  Awschalom.
\newblock Guiding diamond spin qubit growth with computational methods.
\newblock \emph{Physical Review Materials}, 8\penalty0 (2):\penalty0 026204,
  2024.
\newblock \doi{10.1103/PhysRevMaterials.8.026204}.

\bibitem[Horn et~al.(2024)Horn, Wicker, Wellisz, Zeledon, Nittala, Heremans,
  Awschalom, and Guha]{horn2024controlled}
Connor~P Horn, Christina Wicker, Antoni Wellisz, Cyrus Zeledon, Pavani
  Vamsi~Krishna Nittala, F~Joseph Heremans, David~D Awschalom, and Supratik
  Guha.
\newblock Controlled spalling of 4h silicon carbide with investigated spin
  coherence for quantum engineering integration.
\newblock \emph{ACS Nano}, 18\penalty0 (45):\penalty0 31381--31389, 2024.
\newblock \doi{10.1021/acsnano.4c10978}.

\bibitem[Budakian et~al.(2024)Budakian, Finkler, Eichler, Poggio, Degen,
  Tabatabaei, Lee, Hammel, Eugene, Taminiau, et~al.]{budakian2024roadmap}
Raffi Budakian, Amit Finkler, Alexander Eichler, Martino Poggio, Christian~L
  Degen, Sahand Tabatabaei, Inhee Lee, P~Chris Hammel, S~Polzik Eugene, Tim~H
  Taminiau, et~al.
\newblock Roadmap on nanoscale magnetic resonance imaging.
\newblock \emph{Nanotechnology}, 35\penalty0 (41):\penalty0 412001, 2024.
\newblock \doi{10.1088/1361-6528/ad4b23}.

\bibitem[Laraoui et~al.(2013)Laraoui, Dolde, Burk, Reinhard, Wrachtrup, and
  Meriles]{laraoui2013high}
Abdelghani Laraoui, Florian Dolde, Christian Burk, Friedemann Reinhard,
  J{\"o}rg Wrachtrup, and Carlos~A Meriles.
\newblock High-resolution correlation spectroscopy of 13c spins near a
  nitrogen-vacancy centre in diamond.
\newblock \emph{Nature Communications}, 4\penalty0 (1):\penalty0 1651, 2013.
\newblock \doi{10.1038/ncomms2685}.

\bibitem[Abobeih et~al.(2019)Abobeih, Randall, Bradley, Bartling, Bakker,
  Degen, Markham, Twitchen, and Taminiau]{abobeih2019atomic}
MH~Abobeih, J~Randall, CE~Bradley, HP~Bartling, MA~Bakker, MJ~Degen, M~Markham,
  DJ~Twitchen, and TH~Taminiau.
\newblock Atomic-scale imaging of a 27-nuclear-spin cluster using a quantum
  sensor.
\newblock \emph{Nature}, 576\penalty0 (7787):\penalty0 411--415, 2019.
\newblock \doi{10.1038/s41586-019-1834-7}.

\bibitem[Jung et~al.(2021)Jung, Abobeih, Yun, Kim, Oh, Henry, Taminiau, and
  Kim]{jung2021deep}
Kyunghoon Jung, MH~Abobeih, Jiwon Yun, Gyeonghun Kim, Hyunseok Oh, Ang Henry,
  TH~Taminiau, and Dohun Kim.
\newblock Deep learning enhanced individual nuclear-spin detection.
\newblock \emph{npj Quantum Information}, 7\penalty0 (1):\penalty0 41, 2021.
\newblock \doi{10.1038/s41534-021-00377-3}.

\bibitem[Varona-Uriarte et~al.(2024)Varona-Uriarte, Munuera-Javaloy,
  Terradillos, Ban, Alvarez-Gila, Garrote, and Casanova]{varona2024automatic}
B~Varona-Uriarte, C~Munuera-Javaloy, E~Terradillos, Y~Ban, A~Alvarez-Gila,
  E~Garrote, and J~Casanova.
\newblock Automatic detection of nuclear spins at arbitrary magnetic fields via
  signal-to-image ai model.
\newblock \emph{Physical Review Letters}, 132\penalty0 (15):\penalty0 150801,
  2024.
\newblock \doi{10.1103/PhysRevLett.132.150801}.

\bibitem[Kura and Ueda(2018)]{kura2018finite}
Naoto Kura and Masahito Ueda.
\newblock Finite-error metrological bounds on multiparameter hamiltonian
  estimation.
\newblock \emph{Physical Review A}, 97\penalty0 (1):\penalty0 012101, 2018.
\newblock \doi{10.1103/PhysRevA.97.012101}.

\bibitem[Yu et~al.(2023)Yu, Sun, Han, and Yuan]{yu2023robust}
Wenjun Yu, Jinzhao Sun, Zeyao Han, and Xiao Yuan.
\newblock Robust and efficient hamiltonian learning.
\newblock \emph{Quantum}, 7:\penalty0 1045, 2023.
\newblock \doi{10.22331/q-2023-06-29-1045}.

\bibitem[Huang et~al.(2023)Huang, Tong, Fang, and Su]{huang2023learning}
Hsin-Yuan Huang, Yu~Tong, Di~Fang, and Yuan Su.
\newblock Learning many-body hamiltonians with heisenberg-limited scaling.
\newblock \emph{Physical Review Letters}, 130\penalty0 (20):\penalty0 200403,
  2023.
\newblock \doi{10.1103/PhysRevLett.130.200403}.

\bibitem[Anshu et~al.(2021)Anshu, Arunachalam, Kuwahara, and
  Soleimanifar]{anshu2021sample}
Anurag Anshu, Srinivasan Arunachalam, Tomotaka Kuwahara, and Mehdi
  Soleimanifar.
\newblock Sample-efficient learning of interacting quantum systems.
\newblock \emph{Nature Physics}, 17\penalty0 (8):\penalty0 931--935, 2021.
\newblock \doi{10.1038/s41567-021-01232-0}.

\bibitem[Green(1995)]{green1995reversible}
Peter~J Green.
\newblock Reversible jump markov chain monte carlo computation and bayesian
  model determination.
\newblock \emph{Biometrika}, 82\penalty0 (4):\penalty0 711--732, 1995.
\newblock \doi{10.1093/biomet/82.4.711}.

\bibitem[Vousden et~al.(2016)Vousden, Farr, and Mandel]{vousden2016dynamic}
WD~Vousden, Will~M Farr, and Ilya Mandel.
\newblock Dynamic temperature selection for parallel tempering in markov chain
  monte carlo simulations.
\newblock \emph{Monthly Notices of the Royal Astronomical Society},
  455\penalty0 (2):\penalty0 1919--1937, 2016.
\newblock \doi{10.1093/mnras/stv2422}.

\bibitem[Denison et~al.(2002)Denison, Holmes, Mallick, and
  Smith]{denison2002bayesian}
David~GT Denison, Christopher~C Holmes, Bani~K Mallick, and Adrian~FM Smith.
\newblock \emph{Bayesian methods for nonlinear classification and regression},
  volume 386.
\newblock John Wiley \& Sons, 2002.

\bibitem[Sanz-Alonso and Al-Ghattas(2024)]{sanz2024first}
Daniel Sanz-Alonso and Omar Al-Ghattas.
\newblock A first course in monte carlo methods.
\newblock \emph{arXiv preprint arXiv:2405.16359}, 2024.
\newblock \doi{10.48550/arXiv.2405.16359}.

\bibitem[Green and Hastie(2009)]{green2009reversible}
Peter~J Green and David~I Hastie.
\newblock Reversible jump mcmc.
\newblock Technical report, University of Bristol, 2009.

\bibitem[Toubiana et~al.(2023)Toubiana, Katz, and Gair]{toubiana2023there}
Alexandre Toubiana, Michael~L Katz, and Jonathan~R Gair.
\newblock Is there an excess of black holes around 20 m\astrosun? optimizing
  the complexity of population models with the use of reversible jump mcmc.
\newblock \emph{Monthly Notices of the Royal Astronomical Society},
  524\penalty0 (4):\penalty0 5844--5853, 2023.
\newblock \doi{10.1093/mnras/stad2215}.

\bibitem[Zevin et~al.(2017)Zevin, Pankow, Rodriguez, Sampson, Chase, Kalogera,
  and Rasio]{zevin2017constraining}
Michael Zevin, Chris Pankow, Carl~L Rodriguez, Laura Sampson, Eve Chase,
  Vassiliki Kalogera, and Frederic~A Rasio.
\newblock Constraining formation models of binary black holes with
  gravitational-wave observations.
\newblock \emph{The Astrophysical Journal}, 846\penalty0 (1):\penalty0 82,
  2017.
\newblock \doi{10.3847/1538-4357/aa8408}.

\bibitem[Zhu and Gibson(2018)]{zhu2018seismic}
Dehan Zhu and Richard Gibson.
\newblock Seismic inversion and uncertainty quantification using
  transdimensional markov chain monte carlo method.
\newblock \emph{Geophysics}, 83\penalty0 (4):\penalty0 R321--R334, 2018.
\newblock \doi{10.1190/geo2016-0594.1}.

\bibitem[Cho et~al.(2018)Cho, Gibson~Jr, and Zhu]{cho2018quasi}
Yongchae Cho, Richard~L Gibson~Jr, and Dehan Zhu.
\newblock Quasi 3d transdimensional markov-chain monte carlo for seismic
  impedance inversion and uncertainty analysis.
\newblock \emph{Interpretation}, 6\penalty0 (3):\penalty0 T613--T624, 2018.
\newblock \doi{10.1190/INT-2017-0136.1}.

\bibitem[Oaks et~al.(2022)Oaks, Wood~Jr, Siler, and
  Brown]{oaks2022generalizing}
Jamie~R Oaks, Perry~L Wood~Jr, Cameron~D Siler, and Rafe~M Brown.
\newblock Generalizing bayesian phylogenetics to infer shared evolutionary
  events.
\newblock \emph{Proceedings of the National Academy of Sciences}, 119\penalty0
  (29):\penalty0 e2121036119, 2022.
\newblock \doi{10.1073/pnas.2121036119}.

\bibitem[Pagel and Meade(2008)]{pagel2008modelling}
Mark Pagel and Andrew Meade.
\newblock Modelling heterotachy in phylogenetic inference by reversible-jump
  markov chain monte carlo.
\newblock \emph{Philosophical Transactions of the Royal Society B: Biological
  Sciences}, 363\penalty0 (1512):\penalty0 3955--3964, 2008.
\newblock \doi{10.1098/rstb.2008.0178}.

\bibitem[Tak{\'a}cs and Iv{\'a}dy(2024)]{takacs2024accurate}
Istv{\'a}n Tak{\'a}cs and Viktor Iv{\'a}dy.
\newblock Accurate hyperfine tensors for solid state quantum applications: case
  of the nv center in diamond.
\newblock \emph{Communications Physics}, 7\penalty0 (1):\penalty0 178, 2024.
\newblock \doi{10.1038/s42005-024-01668-9}.

\bibitem[Van~de Stolpe et~al.(2024)Van~de Stolpe, Kwiatkowski, Bradley,
  Randall, Abobeih, Breitweiser, Bassett, Markham, Twitchen, and
  Taminiau]{van2024mapping}
GL~Van~de Stolpe, DP~Kwiatkowski, CE~Bradley, J~Randall, MH~Abobeih,
  SA~Breitweiser, LC~Bassett, M~Markham, DJ~Twitchen, and TH~Taminiau.
\newblock Mapping a 50-spin-qubit network through correlated sensing.
\newblock \emph{Nature Communications}, 15\penalty0 (1):\penalty0 2006, 2024.
\newblock \doi{10.1038/s41467-024-46075-4}.

\bibitem[Engquist et~al.(2016)Engquist, Froese, and Yang]{engquist2016optimal}
Bjorn Engquist, Brittany~D Froese, and Yunan Yang.
\newblock Optimal transport for seismic full waveform inversion.
\newblock \emph{arXiv preprint arXiv:1602.01540}, 2016.
\newblock \doi{10.48550/arXiv.1602.01540}.

\bibitem[Oh et~al.(2020)Oh, Yun, Abobeih, Jung, Kim, Taminiau, and
  Kim]{oh2020algorithmic}
Hyunseok Oh, Jiwon Yun, MH~Abobeih, Kyung-Hoon Jung, Kiho Kim, TH~Taminiau, and
  Dohun Kim.
\newblock Algorithmic decomposition for efficient multiple nuclear spin
  detection in diamond.
\newblock \emph{Scientific Reports}, 10\penalty0 (1):\penalty0 14884, 2020.
\newblock \doi{10.1038/s41598-020-71339-6}.

\bibitem[Walker et~al.(2006)Walker, P{\'e}rez-Barber{\'\i}a, and
  Marion]{walker2006stochastic}
David~M Walker, F~Javier P{\'e}rez-Barber{\'\i}a, and Glenn Marion.
\newblock Stochastic modelling of ecological processes using hybrid gibbs
  samplers.
\newblock \emph{Ecological Modelling}, 198\penalty0 (1-2):\penalty0 40--52,
  2006.
\newblock \doi{10.1016/j.ecolmodel.2006.04.008}.

\bibitem[Mathew et~al.(2012)Mathew, Bauer, Koistinen, Reetz, L{\'e}on, and
  Sillanp{\"a}{\"a}]{mathew2012bayesian}
Boby Mathew, AM~Bauer, Petri Koistinen, TC~Reetz, Jens L{\'e}on, and
  MJ~Sillanp{\"a}{\"a}.
\newblock Bayesian adaptive markov chain monte carlo estimation of genetic
  parameters.
\newblock \emph{Heredity}, 109\penalty0 (4):\penalty0 235--245, 2012.
\newblock \doi{10.1038/hdy.2012.35}.

\bibitem[Zhang et~al.(2021)Zhang, Li, Chen, and Li]{zhang2021geological}
Jian Zhang, Jingye Li, Xiaohong Chen, and Yuanqiang Li.
\newblock Geological structure-guided hybrid mcmc and bayesian linearized
  inversion methodology.
\newblock \emph{Journal of Petroleum Science and Engineering}, 199:\penalty0
  108296, 2021.
\newblock \doi{10.1016/j.petrol.2020.108296}.

\bibitem[Reuschen et~al.(2021)Reuschen, Jobst, and
  Nowak]{reuschen2021efficient}
Sebastian Reuschen, Fabian Jobst, and Wolfgang Nowak.
\newblock Efficient discretization-independent bayesian inversion of
  high-dimensional multi-gaussian priors using a hybrid mcmc.
\newblock \emph{Water Resources Research}, 57\penalty0 (8):\penalty0
  e2021WR030051, 2021.
\newblock \doi{10.1029/2021WR030051}.

\bibitem[Nguyen et~al.(2024)Nguyen, Potapenko, Demirci, Govind, Bompas, and
  Sandfeld]{nguyen2024efficient}
Binh~Duong Nguyen, Pavlo Potapenko, Aytekin Demirci, Kishan Govind,
  S{\'e}bastien Bompas, and Stefan Sandfeld.
\newblock Efficient surrogate models for materials science simulations: Machine
  learning-based prediction of microstructure properties.
\newblock \emph{Machine Learning with Applications}, 16:\penalty0 100544, 2024.
\newblock \doi{10.1016/j.mlwa.2024.100544}.

\bibitem[Babbar et~al.(2024)Babbar, Ragunathan, Mitra, Dutta, and
  Patra]{babbar2024explainability}
Agrim Babbar, Sriram Ragunathan, Debirupa Mitra, Arnab Dutta, and Tarak~K
  Patra.
\newblock Explainability and extrapolation of machine learning models for
  predicting the glass transition temperature of polymers.
\newblock \emph{Journal of Polymer Science}, 62\penalty0 (6):\penalty0
  1175--1186, 2024.
\newblock \doi{10.1002/pol.20230714}.

\bibitem[Nyshadham et~al.(2019)Nyshadham, Rupp, Bekker, Shapeev, Mueller,
  Rosenbrock, Cs{\'a}nyi, Wingate, and Hart]{nyshadham2019machine}
Chandramouli Nyshadham, Matthias Rupp, Brayden Bekker, Alexander~V Shapeev, Tim
  Mueller, Conrad~W Rosenbrock, G{\'a}bor Cs{\'a}nyi, David~W Wingate, and
  Gus~LW Hart.
\newblock Machine-learned multi-system surrogate models for materials
  prediction.
\newblock \emph{npj Computational Materials}, 5\penalty0 (1):\penalty0 51,
  2019.
\newblock \doi{10.1038/s41524-019-0189-9}.

\bibitem[Verriere et~al.(2022)Verriere, Schunck, Kim, Marevi{\'c}, Quinlan,
  Ngo, Regnier, and Lasseri]{verriere2022building}
Marc Verriere, Nicolas Schunck, Irene Kim, Petar Marevi{\'c}, Kevin Quinlan,
  Michelle~N Ngo, David Regnier, and Raphael~David Lasseri.
\newblock Building surrogate models of nuclear density functional theory with
  gaussian processes and autoencoders.
\newblock \emph{Frontiers in Physics}, 10:\penalty0 1028370, 2022.
\newblock \doi{10.3389/fphy.2022.1028370}.

\bibitem[Roongcharoen et~al.(2024)Roongcharoen, Conter, Sementa, Melani, and
  Fortunelli]{roongcharoen2024machine}
Thantip Roongcharoen, Giorgio Conter, Luca Sementa, Giacomo Melani, and
  Alessandro Fortunelli.
\newblock Machine-learning-accelerated dft conformal sampling of catalytic
  processes.
\newblock \emph{Journal of Chemical Theory and Computation}, 20\penalty0
  (21):\penalty0 9580--9591, 2024.
\newblock \doi{10.1021/acs.jctc.4c00643}.

\bibitem[Chandrasekaran et~al.(2019)Chandrasekaran, Kamal, Batra, Kim, Chen,
  and Ramprasad]{chandrasekaran2019solving}
Anand Chandrasekaran, Deepak Kamal, Rohit Batra, Chiho Kim, Lihua Chen, and
  Rampi Ramprasad.
\newblock Solving the electronic structure problem with machine learning.
\newblock \emph{npj Computational Materials}, 5\penalty0 (1):\penalty0 22,
  2019.
\newblock \doi{10.1038/s41524-019-0162-7}.

\bibitem[Clyde et~al.(2011)Clyde, Ghosh, and Littman]{clyde2011bayesian}
Merlise~A Clyde, Joyee Ghosh, and Michael~L Littman.
\newblock Bayesian adaptive sampling for variable selection and model
  averaging.
\newblock \emph{Journal of Computational and Graphical Statistics}, 20\penalty0
  (1):\penalty0 80--101, 2011.
\newblock \doi{10.1198/jcgs.2010.09049}.

\bibitem[Seni and Elder(2010)]{seni2010ensemble}
Giovanni Seni and John Elder.
\newblock \emph{Ensemble methods in data mining: improving accuracy through
  combining predictions}.
\newblock Morgan \& Claypool Publishers, 2010.
\newblock \doi{10.2200/S00240ED1V01Y200912DMK002}.

\bibitem[B{\"u}hlmann(2011)]{buhlmann2011bagging}
Peter B{\"u}hlmann.
\newblock Bagging, boosting and ensemble methods.
\newblock In \emph{Handbook of computational statistics: Concepts and methods},
  pages 985--1022. Springer, 2011.
\newblock \doi{10.1007/978-3-642-21551-3_33}.

\bibitem[Adrian et~al.(2024)Adrian, Soloff, and Willett]{adrian2024stabilizing}
Melissa Adrian, Jake~A Soloff, and Rebecca Willett.
\newblock Stabilizing black-box model selection with the inflated argmax.
\newblock \emph{arXiv preprint arXiv:2410.18268}, 2024.
\newblock \doi{10.48550/arXiv.2410.18268}.

\bibitem[Li et~al.(2023)Li, Wang, and Yan]{li2023surrogate}
Yongchao Li, Yanyan Wang, and Liang Yan.
\newblock Surrogate modeling for bayesian inverse problems based on
  physics-informed neural networks.
\newblock \emph{Journal of Computational Physics}, 475:\penalty0 111841, 2023.
\newblock \doi{10.1016/j.jcp.2022.111841}.

\bibitem[Lieberman et~al.(2010)Lieberman, Willcox, and
  Ghattas]{lieberman2010parameter}
Chad Lieberman, Karen Willcox, and Omar Ghattas.
\newblock Parameter and state model reduction for large-scale statistical
  inverse problems.
\newblock \emph{SIAM Journal on Scientific Computing}, 32\penalty0
  (5):\penalty0 2523--2542, 2010.
\newblock \doi{10.1137/090775622}.

\bibitem[Hoffman and Gelman(2014)]{hoffman2014no}
Matthew~D Hoffman and Andrew Gelman.
\newblock The no-u-turn sampler: adaptively setting path lengths in hamiltonian
  monte carlo.
\newblock \emph{Journal of Machine Learning Research}, 15\penalty0
  (1):\penalty0 1593--1623, 2014.
\newblock \doi{10.48550/arXiv.1111.4246}.

\bibitem[Poteshman et~al.(2025)Poteshman, Onizhuk, Egerstrom, Mark, Awschalom,
  Heremans, and Galli]{poteshman2025prapplied}
Abigail~N. Poteshman, Mykyta Onizhuk, Christopher Egerstrom, Daniel~P. Mark,
  David~D. Awschalom, F.~Joseph Heremans, and Giulia Galli.
\newblock High-throughput spin-bath characterization of spin defects in
  semiconductors.
\newblock \emph{Phys. Rev. Appl.}, 24:\penalty0 054048, 2025.
\newblock \doi{10.1103/p57x-8kk7}.

\bibitem[Dr{\'e}au et~al.(2012)Dr{\'e}au, Maze, Lesik, Roch, and
  Jacques]{dreau2012high}
A~Dr{\'e}au, J-R Maze, M~Lesik, J-F Roch, and V~Jacques.
\newblock High-resolution spectroscopy of single nv defects coupled with nearby
  13 c nuclear spins in diamond.
\newblock \emph{Physical Review B}, 85\penalty0 (13):\penalty0 134107, 2012.
\newblock \doi{10.1103/PhysRevB.85.134107}.

\end{thebibliography}

\onecolumn\newpage
\appendix

\section{Hyperparameter dependence}
\label{app:hyperparams}

\begin{figure}[ht]
    \centering
    \includegraphics[width=\textwidth]{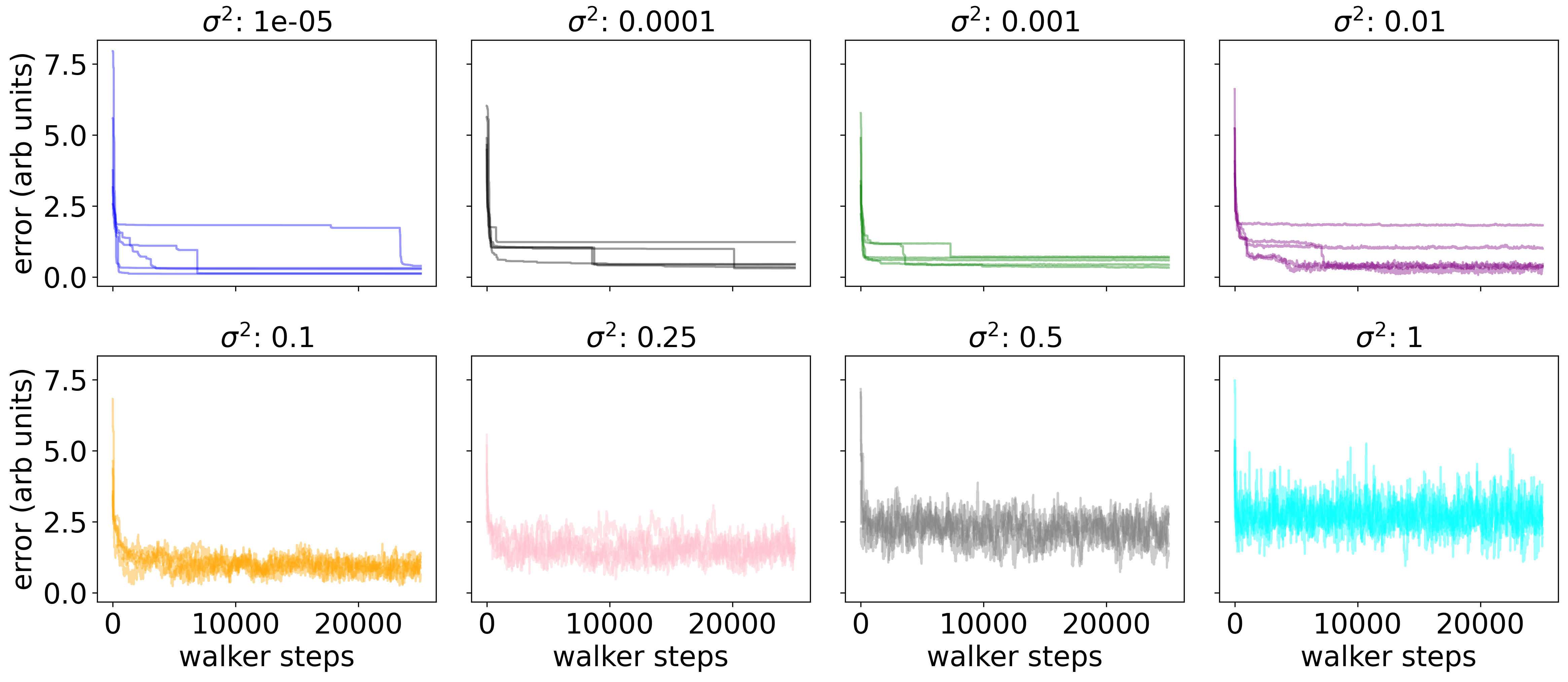} 
    \caption{\textbf{Error trajectories of 5 ensembles as a function of walker steps for varying $\sigma^2$}. The recovery algorithm was applied to the same coherence signal generated using $N=16$ CP pulses, $\varepsilon = 0.001$ shot noise, hyperfine perturbation $0.001$, $\tau_{\text{max}} = 0.008$ ms, and 250 interpulse times sampled from 15 simulated spins with $R_{\text{spin}} = 5$ \AA.  }
    \label{fig:sigma_trials}
\end{figure}

\begin{figure}[ht]
    \centering
    \includegraphics[width=\textwidth]{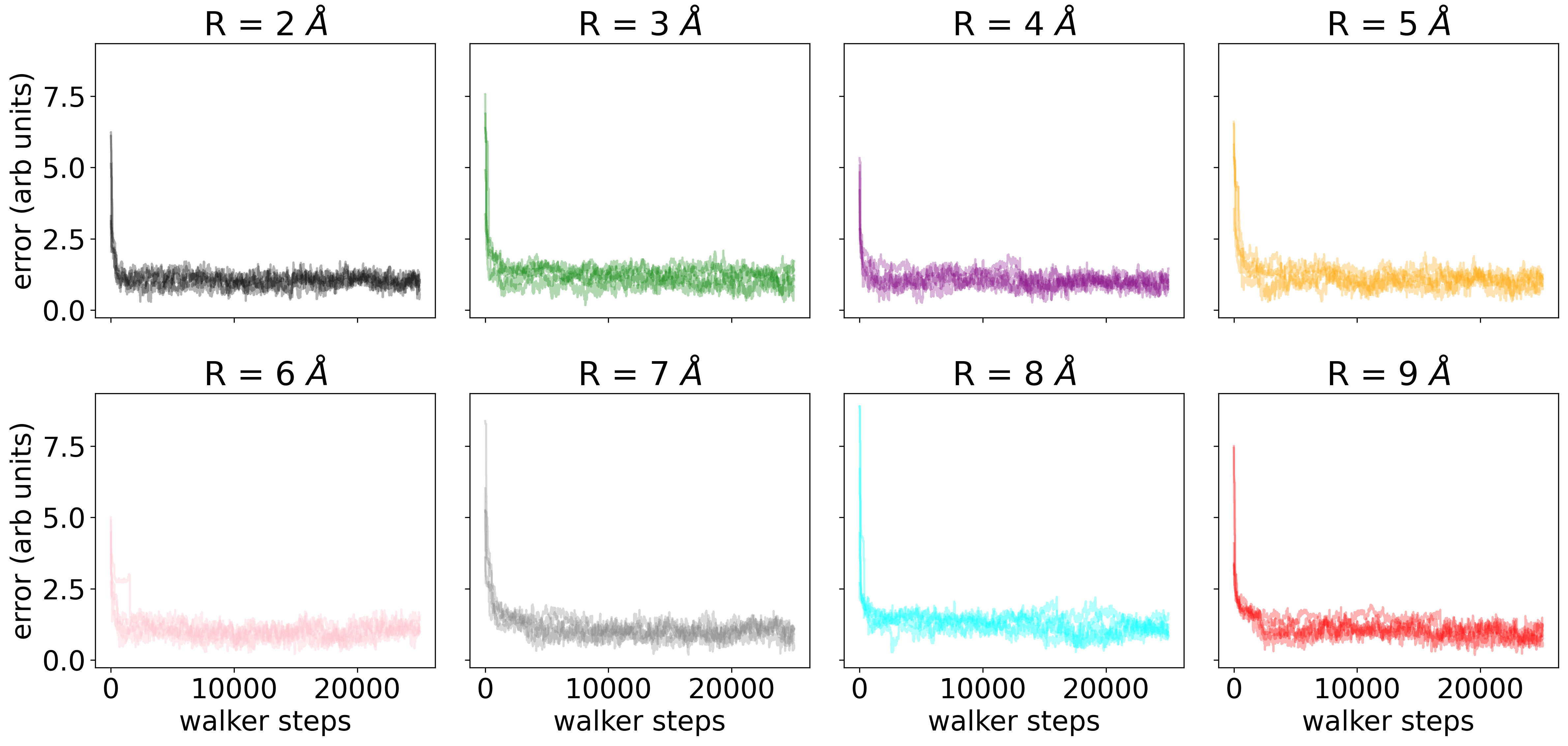} 
    \caption{\textbf{Error trajectories of 5 ensembles as a function of walker steps for varying $R_{\text{spin}}$}. The recovery algorithm was applied to the same coherence signal generated using $N=16$ CP pulses, $\varepsilon = 0.001$ shot noise, hyperfine perturbation $0.001$, $\tau_{\text{max}} = 0.008$ ms, and 250 interpulse times sampled from 15 simulated spins with $\sigma^2 = 0.1$. } 
    \label{fig:R_trials}
\end{figure}

There are a variety of hyperparameters in the hybrid algorithm, including the walker radius of the lattice sites ($R_{\text{spin}}$), the walker radius over $\lambda$ values ($R_{T_2}$), the number of strands for parallel tempering, the inverse temperatures used for parallel tempering, and the noise variance parameter in the likelihood function (Eq. 3 in the main text) $\sigma^2$. While we do not tune every hyperparameter for optimal performance (and we make no claim that the hyperparameters selected for this work are optimal), we note that we get sufficient performance even without having to go through the computationally intensive steps to systematically and exhaustively tune the hyperparameters. 

We find that the hybrid algorithm presented is highly robust to $R_{\text{spin}}$, but the performance of the hybrid algorithm is sensitive to $\sigma^2$. In Fig. \ref{fig:R_trials}, we find that the variance, mean, and convergence properties of the error trajectories are uncorrelated with the size of the walker radius over the lattice sites $R_{\text{spin}}$. That is, the probabilistic selection rules introduced by the hybrid algorithm bias the walkers toward the same solution, regardless of the size of the neighborhood by which each lattice walker is constrained. Notably, there is a physical limit on how small the walker radius can be, which is the distance of nearest neighbors between lattice sites. For diamond, the distance between nearest neighbors is around $1.54$ \AA. Unless otherwise stated, all simulations and experimental data recovery were performed with $R_{\text{spin}} = 5$ \AA.

In contrast, the performance of the hybrid algorithm is highly sensitive to the value of $\sigma^2$ in the likelihood. We find that for values of $\sigma^2$ that are too small, even though the variance in the error is small and the mean errors tend to be low, ensembles are highly likely to be trapped in local minima, whereas for values of $\sigma^2$ that are too large, the mean error and variance both increase (see Fig. \ref{fig:sigma_trials}). Unless stated otherwise, we use $\sigma^2 = 0.1$, which represents the best tradeoff between low variance, low mean, and minimal trapping in local minima.

We note that hyperparameters need not be strictly optimal: beyond a certain point, further optimization yields diminishing returns, and the cost of hyperparameter optimization should not exceed the cost of the inference itself. Our primary requirement for hyperparameter values is reliable convergence of the MCMC walkers to the posterior distribution on an accessible timescale given available computational resources. In practice, tuning the hybrid algorithm involves balancing exploration and convergence across both discrete and continuous parameter spaces, although one should also account for the trade-off between optimizing hyperparameters, which can lead to faster convergence, versus the time-consuming and computationally intensive process of optimizing the hyperparameters. 

For continuous parameters, adjusting proposal step sizes or covariance matrices can help achieve an acceptance rate that promotes efficient mixing while avoiding local trapping. For discrete parameters, such as lattice site selection, the choice of neighborhood radius should respect physical constraints (e.g., nearest-neighbor distances) but can otherwise be broad without degrading convergence. Parallel tempering parameters, including the number of strands and inverse temperatures, can be selected to ensure sufficient exchange between chains, particularly in multimodal or rugged likelihood landscapes. Finally, the noise variance $\sigma^2$ in the likelihood should be tuned to balance model flexibility against overconfidence: too small values risk trapping the walkers in local minima, whereas too large values reduce the algorithm’s ability to accurately recover underlying parameters. In many cases, adaptive strategies provide a practical approach to systematically improve performance without requiring exhaustive hyperparameter searches \cite{hoffman2014no};  such as monitoring acceptance rates and effective sample sizes during burn-in and adjusting proposal distributions or tempering schedules accordingly.

\section{Performance of hybrid algorithm in dense, low noise limit}
\label{app:large_data}

\begin{figure}[ht]
    \centering
    \includegraphics[width=\textwidth]{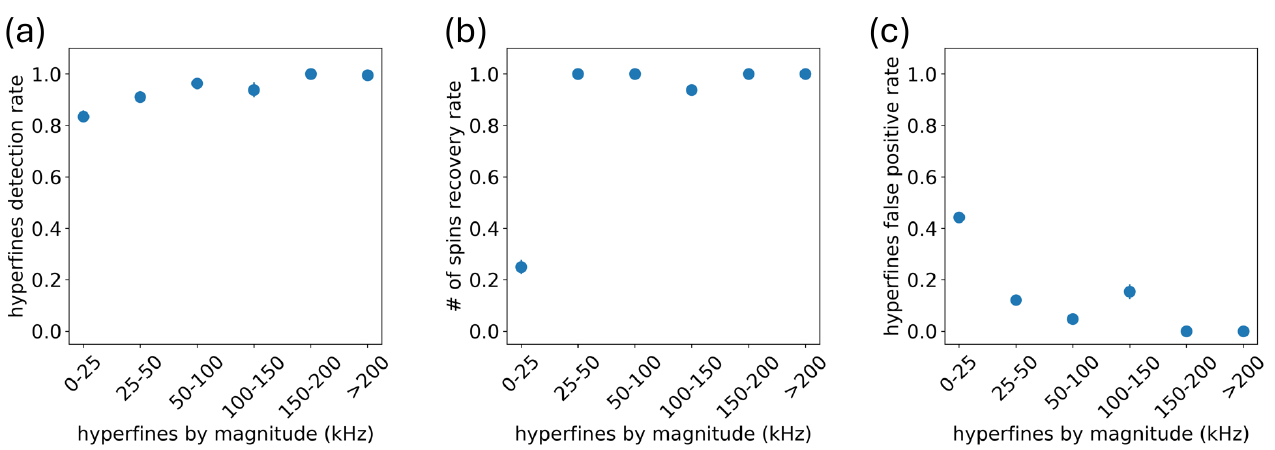} 
    \caption{\textbf{Recovery method on data with experimental parameters in machine learning regime}. \textbf{(a)} Detection rate for hyperfine couplings of the posterior distribution, \textbf{(b)} detection rate for the number of spins in the modal configurations of the posterior distribution, and \textbf{(c)} false positive rate for hyperfine couplings for spins in the modal spin configuration over the posterior distribution for data generated with 15,000 time points for N=32 and N=256 CP pulses at 404 G magnetic field.}
    \label{fig:ml_limit}
\end{figure}

We demonstrate that the proposed method achieves high accuracy, even for weakly coupled spins, with over 80\% accuracy for spins in the 0-25 kHz frequency range when applied to coherence data simulated in the large data limit commonly used in machine learning approaches. When the method is applied to experimental coherence signals that are in the resolution limit of the data used in recent machine learning studies (e.g., \cite{jung2021deep, varona2024automatic}), as shown in Figure \ref{fig:ml_limit}, it achieves low false positive rates and near-perfect accuracy for recovering the number of spins when the coupling frequencies exceed 25 kHz.  

Despite these favorable trends, performance in the 0–25 kHz regime remains substantially degraded in terms of accuracy of the number of spins detected (\~30\%) and false positive rate (\~40\%) as seen in Figure \ref{fig:ml_limit}, even in the limit of long dynamical decoupling sequences, large numbers of dynamical decoupling pulses, and dense samplings of interpulse spacings. This behavior should not be interpreted as an algorithmic failure, but rather as a manifestation of the fundamentally ill-posed nature of the inverse problem associated with weakly coupled nuclear spins. In this regime, individual spins induce only minimal modulations of the NV coherence signal, and the resulting signal variations are often comparable in magnitude to experimental noise \cite{poteshman2025prapplied}. Thus, many distinct spin configurations, differing in both number and hyperfine couplings, produce coherence signals that are effectively indistinguishable within the accessible measurement resolution.

The ill-posedness is further exacerbated by the large number of nuclear spins at small hyperfine couplings, which leads to a combinatorial proliferation of nearly degenerate solutions in parameter space. Even with extensive data, including large numbers of sampled time points and high pulse numbers, the available measurements do not uniquely constrain the underlying spin configuration. In this setting, additional data primarily reinforce existing ambiguities rather than resolve them, reflecting a saturation of the information content of the coherence signal under fixed experimental parameters.

Within our Bayesian framework, this intrinsic non-identifiability is made explicit through the structure of the posterior distribution, which exhibits broad, multi-modal support in the weak-coupling regime. The strict detection criteria employed here, requiring consistency between detected spins and lattice-resolved hyperfine couplings, favor conservative reconstructions at the expense of reduced apparent detection rates for extremely weakly coupled spins. Importantly, this trade-off reflects a principled response to an underdetermined inverse problem rather than a limitation of the inference procedure itself.

These observations indicate that improving performance in the 0–25 kHz regime cannot be achieved solely through algorithmic refinement or increased data density under otherwise fixed experimental conditions. Instead, resolving such weakly coupled spins requires additional, complementary experimental information that alters the structure of the inverse problem itself, for example by combining measurements performed under different magnetic fields. Hence, the reduced performance in the weak-coupling regime reflects a fundamental information limit of the measurement setting, rather than a deficiency of the Bayesian inference framework.

\section{Experimental procedure}
\label{app:exp_procedure}
CPMG data were measured on a single, naturally occurring NV center in a high-purity chemical-vapor-deposition homoepitaxial diamond (type IIa, 1.1\% natural abundance of $^{13}$C) with $\langle 111 \rangle$ crystal orientation. A solid immersion lens with an aluminum-oxide anti-reflection coating (atomic-layer deposition) was used for enhanced photon collection. Microwave fields for electron spin manipulation were delivered via on-chip lithographically-defined strip lines, and a static magnetic field of $B_z \approx 403$ G was applied along the NV-axis using a room-temperature neodymium magnet, with stability $<3$ mG and alignment within $0.07^\circ$ (thermal echo calibration). The electron spin Rabi frequency was $14.31(3)$ MHz, with Hermite pulse shapes for efficient MW pulses, and phase alternation of $\pi$-pulses was performed using the XY-8 scheme. Experiments were conducted at $3.7$ K in a commercial closed-cycle cryostat (Montana Cryostation), achieving single-shot NV spin-state readout fidelity of $94.5\%$ via spin-selective resonant excitation. The electron spin relaxation time was $T_1 > 1$ h, natural dephasing time $T_2 = 4.9(2)$ $\mu$s, spin-echo coherence time $T_2 = 1.182(5)$ ms, and multipulse dynamical decoupling coherence time $T_{2,\text{DD}} > 1$ s with optimized inter-pulse delay $2\tau$.

\section{Dependence on hyperfine couplings}
\label{app:dft_hfs}

\begin{figure}[ht]
    \centering
    \includegraphics[width=\textwidth]{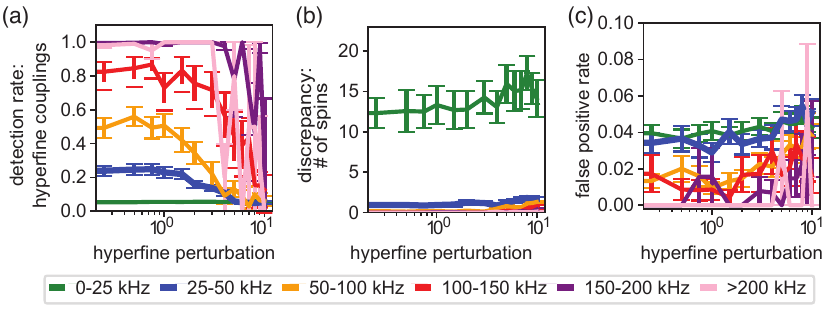} 
    \caption{\textbf{Recovery method accuracy versus hyperfine coupling perturbations}. \textbf{(a)} Detection rate for hyperfine couplings of the posterior distribution, \textbf{(b)} discrepancy between the number of spins recovered and the number of spins simulated, and \textbf{(c)} false positive rate for hyperfine coupling perturbations of different magnitudes. The recovery algorithm was applied to the same coherence signal generated using $N=16$ CP pulses, $\varepsilon = 0.001$ shot noise, $\tau_{\text{max}} = 0.008$ ms, and 250 interpulse times and $\sigma^2 = 0.1$ averaged over simulated nuclear spin baths ranging from 6 to 20 spins.}
    \label{fig:hf_detection}
\end{figure}

Because our model currently constrains the possible hyperfine couplings to those computed from \textit{ab initio} density functional theory for each lattice site, it is essential to understand how the recovery method performs when these input couplings are inaccurate. We investigated the hybrid algorithm's robustness to perturbations in hyperfine couplings, focusing on the ability to recover the hyperfine parameters $A_{\perp}$ and $A_{\parallel}$ from noisy input hyperfine data associated with each lattice site. Simulated inaccuracies in hyperfine couplings, introduced by perturbations $A_{\perp} \pm \delta$ and $A_{\parallel} \pm \delta$, were considered, where we varied the perturbation (denoted as $\delta$) to assess recovery accuracy. We applied our hybrid recovery method to simulations with a  0.001 shot noise per data point, 250 $\tau$ sampled, a 0.008 ms longest $\tau$, 16 CP pulses, and 311 G magnetic field.

Our results indicate that the algorithm remains robust to hyperfine perturbations up to approximately 1 kHz. When the input hyperfine couplings were accurate to the closest kHz, the algorithm performed well, with recovery rates close to 1 for the highest magnitude of hyperfine couplings ($> 150$ kHz). However, when the perturbations exceed 1 kHz, the recovery rates for both hyperfine couplings and number of spins decreases, suggesting that the method is less effective when hyperfine couplings are known with low precision (Fig. \ref{fig:hf_detection}). This threshold suggests that the method is particularly suited for systems where hyperfine couplings can be known to within a kilohertz. Given that state-of-the-art \textit{ab initio} calculations based on density functional theory can achieve relative mean errors of around 1\% for hyperfine couplings at all distances for an NV center in diamond \cite{takacs2024accurate}, and that we seek to detect nuclear spins in the weakly coupled regime (up to 200 kHz), with the majority of spins considered here (over 99\%) below 100 kHz, this level of accuracy falls within the algorithm’s robustness threshold and is promising for practical applications.

However, based on these results for more strongly coupled spins (between 100-500 kHz), we expect a large drop off in accuracy for detection rates for individual spins, since the relative mean error results in a much larger absolute error for more strongly coupled spins. A 1\% mean absolute error for a 250 kHz results in a 2.5 kHz hyperfine perturbation, which is in the regime where the detection rate is much lower than for hyperfine perturbations below 1 kHz and smaller. Nevertheless, the discrepancy between the number of spins for simulated groups of strongly coupled nuclear spins and the recovered number of spins remains relatively low and constant, even for simulations with highly perturbed hyperfine couplings. Taken together, these simulations indicate that this hybrid MCMC approach remains a powerful tool for screening individual samples based on the number of spins of varying magnitudes, even as we expect a decrease in detection accuracy when using theoretical values for nuclear spins with more strongly coupled hyperfine couplings.

\section{Validation of Wasserstein Weighting Parameter \texorpdfstring{$\zeta$}{zeta}}
\label{app:validate_zeta}

To assess the effect of including a Wasserstein distance penalty in the likelihood, we validated the performance of the hybrid MCMC algorithm on experimental data across a range of values for the weighting parameter $\zeta$ in the modified likelihood:
\begin{align}
\mathcal{L}_{\text{mod}}(\mathbf{d}_d | \theta) = 
(1 - \zeta) \cdot \exp\left(
-\frac{1}{2 \sigma^2} \sum_{i=1}^d \left(\mathbf{d}_i - f_k(\mathcal{H}_k(\theta), \tau_i)\right)^2 
\right)
- \zeta \cdot W\left(f_k(\mathcal{H}_k(\theta)), \mathbf{d}\right)
\end{align}
where \( W(\cdot, \cdot) \) denotes the Wasserstein distance between the predicted signal and observed data, and \( \theta = (\mathbf{A_\parallel}, \mathbf{A_\perp}, B_z, \lambda, N) \) are the inferred parameters. In practice, we use the log likelihood to mitigate numerical instability.

\begin{table}[h!]
\centering
\caption{Validation of Wasserstein weighting parameter $\zeta$ on experimental data. The number of nuclear spins detected from the posterior distribution is shown for various values of $\zeta$.}
\label{tab:zeta_validation}
\begin{tabular}{c|c}
\toprule
$\zeta$ & Number of detected spins \\
\hline
\midrule
0     & 20 \\
0.25  & 18 \\
0.5   & 22 \\
0.75  & 12 \\
1.0   & 2  \\
\bottomrule
\end{tabular}
\end{table}

As shown in Table~\ref{tab:zeta_validation}, the value $\zeta = 0.5$ yielded the best balance between signal fidelity and structural regularization, resulting in the recovery of the most physically plausible number of spins (22). At $\zeta = 0$, the inference relies purely on the pointwise squared error, which can overfit noise and result in over-detection. Conversely, large values of $\zeta$ (e.g., $\zeta = 1$) heavily penalize small-scale mismatch and underfit the data, suppressing recovery. 

This weighting is not needed for simulations using ground-truth hyperfine parameters, since both data generation and inference use consistent models (i.e., are drawn from the same distribution). For experimental data, however, uncertainties in measured hyperfine couplings necessitate a regularization term to improve robustness and mitigate overfitting to systematic discrepancies.

\section{Detection rates for hyperfine couplings extracted from experimental data}
\label{app:detect_rates}

We present the detection rates for hyperfine couplings (Table \ref{tab:hf_compare}) and spatial positions (Table \ref{tab:positions_compare}) of nuclear spins, computed from the posterior distribution sampled using the hybrid MCMC algorithm applied to 250 data points from an $N = 32$ CPMG experiment with $\tau$ ranging from 6 to 8 $\mu$s. These results are compared with experimentally reported values in Refs. \cite{abobeih2019atomic, van2024mapping}. We align the computed lattice sites and hyperfine couplings, which were computed in Ref. \cite{takacs2024accurate}, using the HSE06 functional with a supercell containing 1728 atoms, with the experimentally reported values in Ref. \cite{van2024mapping}. If an experimental nuclear spin position is reported with low precision, meaning that multiple lattice sites fall within the range defined by the standard deviation, we identify the ``recovered" lattice site as the single lattice position within this range with the effective spin frequency ($A_{\text{rec}}^-$) that most closely matches the experimentally determined effective spin frequency ($A_{\text{exp}}^-$). There are two nuclear spins, C37 and C49, for which we could not find a good match between the experimentally reported positions and effective spin frequencies with the computed positions and effective spin frequencies. Similarly, while the posterior distribution contains additional hyperfine values corresponding to lattice sites not matched to experimental spins, only the subset of posterior hyperfines that can be associated with experimentally observed spins are reported in Table \ref{tab:hf_compare}.

As noted in Ref. \cite{takacs2024accurate}, there is a known sign inconsistency between the experimentally reported parallel components of hyperfine values and the values computed theoretically. To maintain consistency, we use the sign convention of the experimental data. The hyperfine components computed with the HSE06 hybrid functional in Ref. \cite{takacs2024accurate}, which we also use in this work, were validated against three separate experimental datasets: Refs. \cite{dreau2012high, taminiau2012detection, van2024mapping}. One of these datasets is the same as the one we use for validation here. Because the HSE06 functional was tested on multiple datasets, we consider it a reliable choice for this study and are not overly concerned about overfitting to a single dataset. In Fig. \ref{fig:validation}(d), the yellow points correspond to all posterior hyperfines sampled by the MCMC algorithm, whereas the crosses indicate experimentally observed spins. The crosses plotted in Fig. \ref{fig:validation}(d) correspond to columns 7 and 8 (rather than columns 4 and 5) in Table \ref{tab:hf_compare}, which was chosen to better visualize the correspondence between spins that could be recovered and the posterior hyperfine distribution. That is, all plotted crosses correspond to DFT-computed hyperfines, and in Tables 2 and 3 we provide the subset of spins for which we consider the DFT-computed hyperfines and experimentally measured hyperfines to be equivalent. The matching criterion uses both lattice positions and hyperfine couplings to identify corresponding spins.

We note that the 50 spins associated with this NV center were characterized using different types of experimental measurements, and therefore different amounts of hyperfine information are available across the set. All spins have reconstructed lattice coordinates (Table 3), which we use as the primary criterion to associate experimentally observed spins with DFT lattice sites for hyperfine comparison. For spins 1–27, both $m_s=\pm1$ transition frequencies were measured, allowing direct extraction of $A_{\parallel}$ and $A_{\perp}$ (Table 2). For more information on the experimental procedures for spins 1-27, see \cite{abobeih2019atomic}. For spins 28–50, only partial spectroscopic data are available (the $m_s=-1$ frequency and, for some spins, a local shift $\Delta_i$), which provides information for $A_{\parallel}$ based on the local shift ($\Delta_i \approx A_{\parallel}$), but not for $A_{\perp}$. For more information on the experimental procedures for spins 28-50, see \cite{van2024mapping}, and see the caption of Table \ref{tab:hf_compare} for more details on which approximations were used to compute hyperfine couplings for specific spins. This variation in available experimental data highlights the practical motivation for inference methods that reduce the need for time-consuming full hyperfine characterization measurements.

We calculate the baseline probabilities of nuclear spins appearing in a random selection of 50 spins using the hypergeometric probability distribution. Specifically, we determine the probability that a spin from the set of 3,518 considered lattice sites is included in a random group of 50 spins without replacement, while taking into account the number of symmetrically equivalent lattice sites for each spin with respect to its hyperfine couplings. This approach allows us to compare the observed detection rate of each spin in the posterior distribution with the probability of its random occurrence.

Using the hypergeometric distribution, we determine baseline probabilities of 0.0420 for a site with three symmetrically equivalent sites, 0.0824 for six, 0.1210 for nine, and 0.1581 for twelve. For nuclear spins with identical hyperfine couplings (C8 and C50, C17 and C27, and C42 and C45), we calculate the baseline probability for a second spin based on the probability of a spin appearing at least twice. This results in probabilities of 0.0029 for a site with six symmetrically equivalent sites and 0.0119 for a site with twelve. Using these baseline thresholds, we find that 22 of the nuclear spins reported in \cite{van2024mapping} appear in the posterior distributions sampled by the hybrid MCMC method more frequently than expected by chance.

Some recovered spins have detection rates of zero (see Table \ref{tab:hf_compare}), meaning they are absent in all posterior samples. This occurs in the presence of sparse, noisy experimental data and inaccurate priors, highlighting the ill-posedness of the inverse problem: certain spin configurations cannot be reliably resolved given the limited information content. Importantly, the goal of this Bayesian inference approach is not exact, high-precision reconstruction but rather to provide informative posterior distributions that quantify uncertainty and identify plausible nuclear spin configurations. Hence, the method is particularly useful for rapid, high-throughput screening of spin environments, guiding experimental focus and helping prioritize spins for further investigation, rather than producing perfect reconstructions from limited data.

We do not provide explicit guidelines for a minimum detection rate required to consider a prediction “reliable,” because this threshold is highly context dependent. It can vary based on the specific experimental system, the quality and quantity of available data, and the application or target observable of interest. For example, in regimes with sparse or noisy data, even a moderate detection rate may provide useful information for screening or identifying likely spin candidates, whereas in high-precision applications, only spins with very high detection rates may be considered reliable. Similarly, the acceptable threshold may depend on whether one is interested in qualitative identification of spin presence, quantitative recovery of hyperfine parameters, or downstream tasks such as simulating spin-bath dynamics. In practice, the detection rate should be interpreted alongside other metrics, such as the baseline probability, posterior uncertainty, and consistency with prior knowledge, to assess the confidence in a given prediction.

\begin{longtable}[c]{ c | c c | c c | c c | c}
\hline   
 & $A_{\text{exp}}^-$ & $A_{\text{rec}}^-$ & $A_{\parallel, \text{exp}}$ & $A_{\perp, \text{exp}} $ & $A_{\parallel, \text{rec}}$  & $A_{\perp, \text{rec}} $  & detection\\
 & (kHz) & (kHz) & (kHz) & (kHz) & (kHz) & (kHz) & rate \\
 \hline
 C1 & 452.83(2) & 452.34 & -20.72(1) & 12(1) & -20.65 & 11.18 & 0.0797 \\
C2* & 455.37(2) & 455.01 & -23.22(1) & 13(1) & -23.27 &  13.16 & 0.8760 \\
C3* & 463.27(2) & 462.73 & -31.25(2) & 8(2) & -31.14 & 6.38 &  0.0498\\
C4* & 446.23(4) & 445.73 & -14.07(2) & 13(1) & -14.01 & 12.42  &   0.0564\\
C5 & 447.234(1) & 446.31 & -11.346(2) & 59.21(3) & -10.81 & 59.27 & 0.0583\\
C6 & 480.625(1) & 480.08 & -48.58(2) & 9(2) & -48.39 & 11.51 & 0.0383\\
C7 & 440.288(6) & 439.88 & -8.32(1) & 3(5) & -8.33 & 1.08 & 0.0163\\
C8 & 441.77(1) & 441.37 & -9.79(2) & 5(4) & -9.79 & 4.51 & 0.0683\\
C9 & 218.828(1) & 207.39 & 213.154(1) & 3.0(4) & 224.18 & 3.06 & 0.0603\\
C10 & 414.407(1) & 413.76 & 17.643(1) & 8.6(2) & 17.88 & 8.75 & 0.0679\\
C11* & 417.523(4) & 416.54 & 14.548(3) & 10(1) & 15.13 & 9.85  & 0.1585 \\
C12 & 413.477(1) & 412.07 & 20.569(1) & 41.51(3) & 21.57 & 41.50 & 0.0036\\
C13 & 424.499(1) & 424.15 & 8.029(1) & 21.0(4) & 7.92 & 21.04 &  0.0 \\
C14 & 451.802(1) & 451.38 & -19.815(3) & 5.3(5) & -19.79 & 6.19 & 0.0593\\
C15* & 446.01(5) & 445.71 & -13.961(3) & 9(1) & -14.07 & 9.43 &  0.1108\\
C16* & 436.67(5) & 436.26 & -4.66(3) & 7(4) & -4.65 & 7.44 &  0.1025\\
C17* & 437.61(1) & 437.24 & -5.62(1) & 5(2) & -5.68 & 1.33 & 0.0991\\
C18 & 469.02(1) & 468.14 & -36.308(1) & 26.62(4) & -35.83 & 26.69 &  0.0 \\
C19 & 408.317(1) & 407.58 & 24.399(1) & 24.81(4) & 24.57 & 22.12 &  0.0035 \\
C20 & 429.403(4) & 429.02 & 2.690(4) & 11(1) & 2.67 & 10.96 & 0.0462\\
C21* & 430.937(3) & 430.58 & 1.212(5) & 13(1) & 1.16 & 12.82 & 0.0966\\
C22* & 424.289(3) & 423.83 & 7.683(4) & 4(3) & 7.75 & 5.37 &  0.1092 \\
C23 & 435.143(7) & 436.19 & -3.177(5) & 2(4) & -4.61 & 5.23 &  0.1276\\
C24* & 436.183(3) & 436.84 & -4.225(4) & 0(6) & -5.29 & 1.04 & 0.1256\\
C25 & 435.829(5) & 436.89 & -3.873(5) & 0(4) & -5.33 &  1.66 &  0.1519 \\
C26 & 435.547(2) & 435.23 & -3.618(5) & 0(2) & -3.58 & 9.59 &  0.0982\\
C27* & 435.99(3) & 437.24 & -4.039(5) & 0(3) & -5.68 &  1.33 & 0.0035\\
C28 & 440.9(1) & 440.51 & -8.91547(3)$^{\dagger}$ & 4.65$^{\dagger}$ & -8.48 & 5.90 & 0.0287\\
C29* & 434.3(1) & 433.83 & -2.1857(1)$^{\dagger}$ & 14.86$^{\dagger}$ & -2.09 & 13.13 & 0.0890\\
C30 & 427.1(1) & 426.64 & 4.87111(4)$^{\dagger}$ & 3.08$^{\dagger}$ & 4.94 & 5.43 &  0.1100 \\
C31 & 428.3(1) & 423.09 & -3.66$^{\ddagger}$ & 0$^{\ddagger}$ & 8.48 & 4.32 &  0.0538 \\
C32* & 431.6(1) & 415.60 & -0.36$^{\ddagger}$ &  0$^{\ddagger}$ & 15.96 & 2.36 & 0.0835 \\
C33* & 439.0(1) & 438.58 & 7.04$^{\ddagger}$&  0$^{\ddagger}$ & -7.00 & 4.76 & 0.1137\\
C34 & 437.3(1) & 436.88 & 5.34$^{\ddagger}$ &  0$^{\ddagger}$ & -5.32 & 3.12 & 0.0703 \\
C35 & 427.4(1) & 424.82 & 4.59133(4)$^{\dagger}$ & 5.17$^{\dagger}$ & 6.74 & 3.33 & 0.1213 \\
C36 & 434.4(1) & 365.41 & -2.2142(8)$^{\dagger}$ & 14.00$^{\dagger}$ & 84.52 & 114.42 & 0.0 \\
C37 & 429.1(1) & - & -2.86$^{\ddagger}$ & 0$^{\ddagger}$ & - & - &  - \\
C38* & 434.0(1) & 433.44 & 2.04$^{\ddagger}$ &  0$^{\ddagger}$ & -1.83 & 7.03 & 0.0996 \\
C39 & 432.5(1)& 431.38 & -0.450(5)$^{\dagger}$ & 8.82$^{\dagger}$ & 0.20 & 5.15 &  0.0267 \\
C40* & 433.3(1) & 433.24 & -1.173(5)$^{\dagger}$ & 12.03$^{\dagger}$ & -1.66 & 5.75 & 0.0979 \\
C41* & 434.1(1) & 430.69 & -2.1893(6)$^{\dagger}$ & 0$^{\dagger}$& 0.90 & 6.15 & 0.1749 \\
C42 & 434.8(1) & 437.08 & 2.84$^{\ddagger}$ &  0$^{\ddagger}$ & -5.52 & 3.58 & 0.1462\\
C43* & 432.2(1) & 427.83 & -0.2701(6)$^{\dagger}$ & 0$^{\dagger}$ & 3.94 & 13.63 & 0.0904 \\
C44 & 433.9(1) & 435.32 & -1.8828(6)$^{\dagger}$ & 7.05$^{\dagger}$ & -3.74 & 5.46 &  0.0903\\
C45* & 436.2(1) & 437.08 & -4.174(1)$^{\dagger}$ & 7.59$^{\dagger}$ & -5.52 & 3.58 & 0.0176\\
C46 & 434.8(1) & 430.43 & 2.84$^{\ddagger}$ &  0$^{\ddagger}$ & 1.15 & 5.31 &  0.0332 \\
C47* & 429.4(1) & 423.98 & 2.5878(3)$^{\dagger}$ & 4.89$^{\dagger}$& 7.51 & 2.56 & 0.1065 \\
C48* & 431.0(1) & 394.41 & -0.96$^{\ddagger}$ &  0$^{\ddagger}$ & 37.51 & 17.04 & 0.9588\\
C49 & 428.3(1) & - & -3.66$^{\ddagger}$ & 0$^{\ddagger}$ & - & - & - \\
C50 & 436.2(1) & 207.39 & -4.2270(4)$^{\dagger}$ & 3.37$^{\dagger}$ & 224.19 & 3.06 &  0.0026\\
\hline
\caption{\textbf{Comparison of Experimentally Measured and Recovered Hyperfine Interaction Values.} 
This table compares experimentally measured and recovered hyperfine interaction values for nuclear spins, following the nuclear spin labeling convention in \cite{abobeih2019atomic, van2024mapping}. The experimental data in this Table combines data from two different kinds of experimental measurements. The data for nuclear spins C1-C27 is from \cite{abobeih2019atomic}, in which the hyperfine components ($A_{\parallel, \text{exp}}, A_{\perp,\text{exp}}$) are fit from direct measurements of the $^{13}$C spin precession frequencies for the $m_s = \pm 1$ electron spin projections (see Sec. I of the SI of \cite{abobeih2019atomic} for more information about the experimental measurements and the fitting procedures). The data for nuclear spins C28-C50 is from \cite{van2024mapping}, in which only the spin frequencies $(A^{-}_{\text{exp}})$ corresponding to the $m_s = -1$  electron projection are measured, and for a subset of these nuclear spins (denoted by $^{\dagger}$), a hyperfine shift $\Delta_i$ is measured (see Supplementary Note I of \cite{van2024mapping} for more details on experimental measurements and a derivation of the following approximations). For the nuclear spins denoted by $^{\dagger}$, we approximate $A_{\parallel, \text{exp}} \approx \Delta_i$, and we recover $A_{\perp, \text{exp}}$ based on $A^{-}_{\text{exp}} = \sqrt{(\omega_L - A_{\parallel, \text{exp}})^2 + A_{\perp, \text{exp}}^2}$, and any imaginary numbers are reported as 0$^{\dagger}$. For the spins (denoted with $^{\ddagger}$) where $\Delta_i$ was not measured (due to low polarization, coherence, or readout contrast), we approximate $A_{\parallel, \text{exp}} \approx \omega_L - A^{-}_{\text{exp}}$, where we have assumed that $A_{\parallel, \text{exp}}$ is small and $A_{\perp, \text{exp}} \approx 0$. These experimental values are compared with the hyperfine components ($A_{\parallel, \text{rec}}, A_{\perp, \text{rec}}$) recovered using a hybrid MCMC algorithm (see Sec. \ref{subsec:validation}). We report $A^{-}_{\text{rec}}$ as $A^{-}_{\text{rec}} = \sqrt{(\omega_L - A_{\parallel, \text{rec}})^2 + A_{\perp, \text{rec}}^2}$. Detection rates of individual nuclear spins are determined based on the frequency of their appearance in spin configurations sampled from the posterior distribution. Spins can only be identified up to a degree of symmetry determined by the number of lattice sites with identical hyperfine couplings; however, we properly account for the multiplicity of spins with identical hyperfine couplings but different lattice sites when computing detection rates. Spins with detection rates above their baseline probabilities are marked with an asterisk (*).}
\label{tab:hf_compare}
\end{longtable}

\begin{longtable}[c]{c | c c c | c c c | c}
\hline
& \multicolumn{3}{c|}{Experimental values} & \multicolumn{3}{c|}{Recovered values} & \# of\\
 & $x$ & $y$ & $z$ & $x$ & $y$ & $z$ & symmetric  \\
 & (\AA) & (\AA) & (\AA) & (\AA) & (\AA) & (\AA) & spins \\
 \hline
 C1 & 0.0 & 0.0 & 0.0 & 0.0 & 0.0 & 0.0 & 6\\
 C2* & 2.52 & 2.91 & -0.51 & 2.52 & 2.91 & -0.52 & 6\\
 C3* & 3.78 & 0.73 & -0.51 & 3.78 & 0.73 & -0.52 & 3\\
 C4* & -1.26 & 2.18 & 0.0 & -1.26 & 2.18 & -0.01 & 3\\
 C5 & 0.0 & 4.37 & -6.18 & 0.0 & 4.36 & -6.18 & 6\\
 C6 & 5.04 & -1.46 & -2.06 & 5.04 & -1.46 & -2.06 & 3\\
 C7 & 5.04 & -1.46 & 5.66 & 5.04 & -1.46 & 5.66  & 3\\
 C8 & 7.57 & 1.46 & 3.6 & 7.56 & 1.45 & 3.60 & 6\\
 C9 & 7.57 & -4.37 & -10.81 & 7.56 & -4.37 & -10.82 & 6\\
 C10 & 0.0 & 8.74 & -12.36 & 0.0 & 8.73 & -12.36 & 6\\
 C11* & 6.31 & 9.46 & -12.87 & 6.30 & 9.46 & -12.88 & 6 \\
 C12 & 11.35 & 0.73 & -14.42 & 11.35 & 0.73 & -14.42 & 6 \\
 C13 & 12.61 & 2.91 & -6.69 & 12.61 & 2.91 & -6.70 & 6\\
 C14 & 5.04 & -2.91 & -22.65 & 5.04 & -2.91 & -22.66 & 6\\
 C15* & 1.26 & 3.64 & -22.65 & 1.26 & 3.64 & -22.66 & 6\\
 C16* & 2.52 & 8.74 & -23.17 & 2.52 & 8.74 & -23.17 & 6\\
 C17* & 6.31 & -2.18 & -29.34 & 6.30 & -2.19 & -29.35 & 6\\
 C18 & 0.0 & -1.46 & -19.05 & 0.0 & -1.46 & -19.06 & 6\\
 C19 & 3.78 & -9.46 & -8.75 & 3.78 & -9.46 & -8.76 & 3\\
 C20* & 3.78 & 10.92 & -4.63 & 3.78 & 10.92 & -4.64 & 3\\
 C21* & -5.04 & 5.82 & -4.12 & -5.05 & 5.82 & -4.12 & 6\\
 C22* & 16.39 & -3.64 & -8.24 & 16.39 & -3.64 & -8.24 & 6\\
 C23 & 13.88 & 0.73 & 5.66 & 12.61 & 1.45 & 3.60 & 12\\
 C24* & 1.26 & -0.73 & 9.78 & 1.26 & -0.73 & 8.23 & 6\\
 C25 & 7.57 & 1.46 & 9.78 & 6.30 & 2.18 & 7.72 & 12\\
 C26 & 12.61 & -5.82 & -0.51 & 12.61 & -5.83 & -0.52 & 9\\
 C27* & 1.26 & 3.64 & -31.4 & 1.26 & 2.18 & -29.35 & 6\\
 C28 & -1.26 & 2.18 & -24.71 & -1.26 & 2.18 & -24.72 & 3\\
 C29* & -6.31 & 0.72 & -19.05 & -6.31 & 0.73 & -19.05 & 6\\
 C30 & 12.62 & 10.19 & -14.93 & 12.61 & 10.19 & -10.30 & 12\\
 C31 & 11(4) & 15(4) & -11(4) & 7.56 & 11.65 & -12.88 & 6\\
 C32* & 6(3) & 12(2) & -3(9) & 5.04 & 10.19 & -14.94 & 6\\
 C33* & -2.52 & -1.46 & 4.12 & -2.53 & -1.46 & 4.11 & 6\\
 C34 & -2.52 & 0.0 & 6.18 & -2.53 & 0.0 & 6.17 & 6\\
 C35 & 20(3) & -1(7) & -8(1) & 17.65 & 1.45 & -8.76 & 12\\
 C36 & -8(7) & -0(7) & -23(9) & -1.26 & 0.73 & -14.42 & 6\\
 C37 & 13(2) & -12(1) & -18(1) & - & - & - & -\\
 C38* & 12.61 & -8.73 & 0.0 & 12.61 & -8.74 & -0.01 & 6\\
 C39 & 16(5) & -8(5) & -24(4) & 16.39 & -8.01 & 20.60 & 3\\
 C40* & 11(3) & -14(3) & -26(4) & 10.09 & -11.65 & -22.66 & 6\\
 C41* & 8(4) & -12(4) & 2(6) & 8.82 & -13.84 & -2.58 & 12\\
 C42 & 10.09 & -7.28 & -28.83 & 8.82 & -5.10 & -27.29 & 12\\
 C43* & 11(3) & 7(3) & -19(3)& 12.61 & 5.82 & -16.48 & 6\\
 C44 & 0.0 & 11.65 & -26.77 & 0.0 & 8.74 & -24.72& 9\\
 C45* & 6.3 & 6.56 & -29.35 & 5.04 & 5.82 & -27.29 & 12\\
 C46 & 14(4) & -5(4) & -24(5) & 17.65 & -5.83 & -19.05 & 6\\
 C47* & 21(4) & -7(7) & -14(7) & 17.65 & 0.0 & -12.36 & 6\\
 C48* & 16(4) & -4(4) & -21(9) & 12.61 & 0.0 & -12.36 & 6\\
 C49 & 23(4) & -4(8) & -12(3) & - & - & - & -\\
 C50 & 11(7) & 5(7) & -2(9) & 5.04 & 4.36 & -10.82 & 6\\
\hline
\caption{\textbf{Comparison of experimentally measured and recovered nuclear spin positions.} We present the spatial coordinates ($x$, $y$, $z$) of nuclear spins determined experimentally \cite{abobeih2019atomic, van2024mapping} alongside the recovered coordinates obtained using a hybrid MCMC algorithm. The lattice used in the hybrid MCMC algorithm was aligned to be consistent with the experimentally reported nuclear lattice positions and the experimental values of effective nuclear spin frequencies and hyperfine couplings. The column labeled "\# of symmetric spins" indicates the number of lattice sites with identical hyperfine couplings. Of these symmetric lattice sites, the position closest to the experimentally reported value is shown. }
\label{tab:positions_compare}
\end{longtable}

\end{document}